%% file: hemp2013_integral.tex
\documentclass[iop]{emulateapj}

\usepackage{pslatex}
\usepackage{aas_macros}
\usepackage{amssymb,amsmath,amsxtra,amsfonts}
\usepackage{natbib}
\usepackage{xspace}
\usepackage{hyperref}

\newcommand{\fif}{4U~1538$-$522\xspace}
\newcommand{\nin}{4U~1907+09\xspace}
\newcommand{\msol}{M$_{\odot}$\xspace}
\newcommand{\qvnor}{QV Nor\xspace}
\newcommand{\evslum}{keV/($10^{37}$\,erg\,s$^{-1}$)\xspace}

\accepted{30 August 2013}

\begin{document}

\title{Measurements of Cyclotron Features and Pulse Periods in the High-Mass X-Ray
  Binaries \fif and \nin with \textit{INTEGRAL}}

\author{Paul~B.~Hemphill \altaffilmark{1}, 
Richard~E.~Rothschild \altaffilmark{1},
Isabel~Caballero \altaffilmark{3},
Katja~Pottschmidt \altaffilmark{3,4},
Matthias~K\"uhnel \altaffilmark{5},
Felix~F\"urst \altaffilmark{6},
J\"orn~Wilms \altaffilmark{5}
}

\altaffiltext{1}{Center for Astrophysics and Space Sciences, University of California, San Diego, 9500 Gilman Dr., La Jolla, CA 920093-0424, USA}
\altaffiltext{2}{CEA Saclay, DSM/IRFU/SAp-UMR AIM (7158) CNRS/CEA/Universit\'{e} Paris 7, Diderot, 91191 Gif-sur-Yvette, France}
\altaffiltext{3}{Center for Space Science and Technology, University of Maryland Baltimore County, 1000 Hilltop Circle, Baltimore, MD 21250, USA}
\altaffiltext{4}{CRESST and NASA Goddard Space Flight Center, Astrophysics Science Division, Code 661, Greenbelt, MD 20771, USA}
\altaffiltext{5}{Dr.\ Karl Remeis-Sternwarte \& Erlangen Center for Astroparticle Physics, Sternwartstr. 7, 96049 Bamberg, Germany}
\altaffiltext{6}{Cahill Center for Astronomy and Astrophysics, California Institute of Technology, MC 290-17, 1200 E. California Blvd., Pasadena, CA 91125, USA}

\email{pbhemphill@physics.ucsd.edu}

\begin{abstract}

  We present a spectral and timing analysis of \textit{INTEGRAL} observations of
  two high mass X-ray binaries, \fif and \nin. Our timing measurements for \fif
  find the pulse period to have exhibited a spin-up trend until approximately
  2009, after which there is evidence for a torque reversal, with the source
  beginning to spin down to the most recently-measured period of $525.407 \pm
  0.001$\,s. The most recent \textit{INTEGRAL} observations of \nin are not
  found to yield statistically significant pulse periods due to the
  significantly lower flux from the source compared to \fif. A spectral model
  consisting of a power-law continuum with an exponential cutoff and modified by
  two cyclotron resonance scattering features is found to fit both sources well,
  with the cyclotron scattering features detected at $\sim\!22$ and
  $\sim\!49$\,keV for \fif and at $\sim\!18$ and $\sim\!36$\,keV in \nin. The
  spectral parameters of \fif are generally not found to vary significantly with
  flux, and there is little to no variation across the torque reversal.
  Examining our results in conjunction with previous work, we find no evidence
  for a correlation between cyclotron line energy and luminosity for \fif. \nin
  shows evidence for a positive correlation between cyclotron line energy and
  luminosity, which would make it the fourth, and lowest-luminosity, cyclotron
  line source to exhibit this relationship.
\end{abstract}


\maketitle

\section{Introduction}
\label{sec:intro}

High-mass X-ray binaries (HMXBs) are a class of binary systems consisting of a
neutron star with a high-mass ($\sim\!10-20$ \msol) main-sequence companion.
X-ray emission from the neutron star is driven primarily by accretion of
material from the main-sequence companion, typically via either Roche lobe
overflow in the case of less-massive companions, or the stellar wind of a
higher-mass companion. The neutron star's strong ($\sim\!10^{12-13}$\,G)
magnetic field channels accreted material onto the magnetic poles, forming an
accretion column where the majority of the X-ray luminosity is produced via
inverse Compton scattering of thermal photons produced at the stellar
surface and bremsstrahlung radiation from within the column. Obtaining a
detailed physical picture of this process is still an open problem in X-ray
astrophysics. Early work
\citep[e.g.,][]{meszaros_comptonization_1985b,klein_accretion_1996} was
generally limited in its ability to properly reproduce the accreting neutron
star spectrum, due to the extreme complexity of the problem. Recently, however,
work by \citet{becker_continuum_2007} has produced promising results,
successfully reproducing the continuum spectra of several high-luminosity X-ray
pulsars. In addition to the base continuum, several dozen HMXBs display
absorption-like features in their spectra that are identified as cyclotron
resonance scattering features (CRSFs, also known as cyclotron lines), produced
by the resonant scattering of photons on electrons moving in the magnetic
field; the physical modeling of these features has also seen significant
progress in work by \citet{schoenherr_CRSF_2007}. Work is currently underway to
merge the continuum and cyclotron line models into a single model capable of
producing an accurate, physically-motivated model for accreting neutron stars
\citep{schwarm_headposter_2013,schwarm_thesis_2013}.

The accretion-powered X-ray pulsar \fif was discovered in the early 1970s by the
\textit{Uhuru} satellite \citep{giacconi_third_1974}. Its main-sequence
companion is the $\sim\!20$\,\msol B0 star \qvnor
\citep{reynolds_optical_1538_1992}. X-ray pulsations were first observed from
the vicinity of \fif by \citet{becker_1538_1977} and \citet{davison_1538_1977a};
\citet{davison_1538_1977b} identified these as coming from \fif. The
distance to \fif has been variously estimated to be $5.5 \pm 1.5$
\citep{crampton_1538_1978}, $6.0 \pm 0.5$ \citep{ilovaisky_1538_1979}, $6.4
\pm 1.0$ \citep{reynolds_optical_1538_1992}, and $4.5$
\citep{clark_chandra1538_2004} kpc. Infrequent
observations post-discovery showed \fif to be spinning down until the late
1980s, reaching a maximum period of $\sim\!531$\,s
\citep{makishima_spectra_1987,davison_1538_1977a,becker_1538_1977}, after which
extended BATSE monitoring between 1990 and 1995 by
\citet{rubin_observation_1997} showed the source to have transitioned to a
long-term spin-up trend that would last until approximately 2009, when
monitoring by the Gamma-ray Burst Monitor (GBM) aboard the \textit{Fermi}
satellite, along with this work, show the source to have transitioned to a
spin-down state (see Fig. \ref{fig:H1538_period}). The early \textit{OSO-8}
and \textit{Ariel-5} observations by \citet{becker_1538_1977} and
\citet{davison_1538_1977b} also established a $\sim\!3.7$\,day orbital period
and provided the first picture of the binary orbit of the system, with the
orbital parameters being updated by \citet{makishima_spectra_1987},
\citet{clark_orbit_2000}, and \citet{mukherjee_orbital_2006}. The exact
characteristics of \fif's orbit remain somewhat uncertain: the three
aforementioned references find three different values for the eccentricity,
with \citeauthor{makishima_spectra_1987} adopting $e = 0.08 \pm 0.05$,
\citeauthor{clark_orbit_2000} finding solutions for both circular and
elliptical ($e = 0.174 \pm 0.015$) orbits, and
\citeauthor{mukherjee_orbital_2006} supporting \citeauthor{clark_orbit_2000}'s
elliptical solution. Efforts to determine the mass of \fif by
\citet{vankerkwijk_masses_1995} adopted a circular orbit, while recent work by
\citet{rawls_mass_2011} finds a mass of $0.87 \pm 0.07$\,\msol for
\citeauthor{clark_orbit_2000}'s elliptical orbit and $1.104 \pm 0.177$\,\msol
for the circular solution, making \fif potentially one of the lightest neutron
stars known. The spectrum of \fif is fit well by a power law with an
exponential cutoff \citep{makishima_spectra_1987}, with CRSFs at 22
\citep{clark_discovery_1990} and $\sim\!47$\,keV, tentatively reported in
\textit{BeppoSAX} data by \citet{robba_bepposax_2001}, but not confirmed until
a combined \textit{RXTE} and \textit{INTEGRAL} analysis by
\citet{rodes-roca_first_2009}.

\nin is another wind-accreting X-ray binary, also discovered by \textit{Uhuru}
\citep{giacconi_x-ray_1971}. It is similar to \fif in its long pulse period
($\sim\!440$\,s), short orbital period ($\sim\!8.4$\,d), and optical companion
\citep[O8/O9, as found by][]{cox_1907_2005}. For $\sim\!15$\,yr following the
identification of the source as a pulsar by \citet{makishima_1907_1984}, the
pulse period increased steadily from 437.4\,s in 1984
\citep{makishima_1907_1984} to $\sim\!441$\,s in 1998, when the spin-down trend
began to slow, leading to a torque reversal in 2005 \citep{fritz_1907_2006}.
This was followed relatively quickly by another reversal in mid-2007
\citep{inam_torque_2009,sahiner_recent_2011}, and the source began following
approximately the same spin-down trend as it had between 1987 and 1998, a trend
which thus far shows no sign of changing. The initial determination of the
$\sim\!8.4$\,d orbital period showed flaring twice per orbit, suggesting the
source was in orbit around a Be-type star \citep{marshall_1907_1980}, although
subsequent optical observations indicated that the companion was more likely a
supergiant \citep{vankerkwijk_1907_1989}.  Recent optical
\citep{cox_1907_2005} and infrared \citep{nespoli_1907_2008} observations have
more firmly established the companion as a O8/O9 supergiant, with
\citeauthor{cox_1907_2005} finding a distance of $5.0$\,kpc to the source,
while \citet{kostka_evidence_2010} have shown that the most likely scenario to
produce the observed flaring involves a dense stellar wind along with a
trailing stream of material between the neutron star and companion.
Observations with \textit{Ginga} established the existence of an absorption
feature, identified as a CRSF, at $\sim\!19$\,keV
\citep{mihara_thesis_1995,makishima_1907_1999}, with a second at
$\sim\!39$\,keV detected in \textit{BeppoSAX} data by
\citet{cusumano_1907_1998}.

In this paper, we first present a summary of the observations used in our
analysis in \S \ref{sec:obs}, followed by a brief discussion of the pulse timing
measurements made for each source in \S \ref{sec:timing}. \S \ref{sec:spectral}
presents the spectral analysis of each source, where the large energy range
covered by the combination of the ISGRI and JEM-X instruments proves
advantageous. Finally, we discuss these results in the context of recent
theoretical results regarding accretion onto magnetic poles in \S
\ref{sec:discussion}.

\section{Observations}
\label{sec:obs}

We used two of the instruments aboard \textit{INTEGRAL}: the \textit{INTEGRAL}
Soft Gamma-Ray Imager \citep[ISGRI,][]{lebrun_isgri_2003}, which is the upper
layer of the Imager on Board the \textit{INTEGRAL} Satellite
\citep[IBIS,][]{ubertini_ibis_2003}, a coded-mask telescope with a $9^\circ
\times 9^\circ$ fully-coded field of view, and the twin Joint European X-ray
Monitors \citep[JEM-X 1 and 2,][]{lund_jemx_2003}, coded-mask telescopes with
circular fields of view of diameter $4.8^\circ$. Early in the
\textit{INTEGRAL} mission, only one JEM-X telescope was in operation
at any given time, out of concern for the lifetime of the microstrip anodes
used in the detectors, but lowered operating voltage alleviated these concerns,
and recent (post-2009) observations have both detectors running simultaneously.
The \textit{INTEGRAL} satellite has a $\sim\!3$-day orbit; the science data for
each orbit (or ``revolution'') is divided into $\sim\!2$\,ks-long pointings
called Science Windows (SCWs). The nominal energy range of ISGRI is 15\,keV -
1\,MeV, but we follow the ISGRI team's recommendation of a lower bound of
18\,keV for data taken prior to revolution 848, and 20\,keV for data prior to
revolution 1000. The JEM-X data are limited to $\ge 5$\,keV for similar
reasons. Neither source is detected above $\sim\!80$\,keV in ISGRI or
$\sim\!30$\,keV in JEM-X. We extracted spectra and lightcurves using version
10.0 of the standard Offline Scientific Analysis (OSA) software provided by the
\textit{INTEGRAL} Science Data Center (ISDC).

\begin{figure*}
  \plottwo{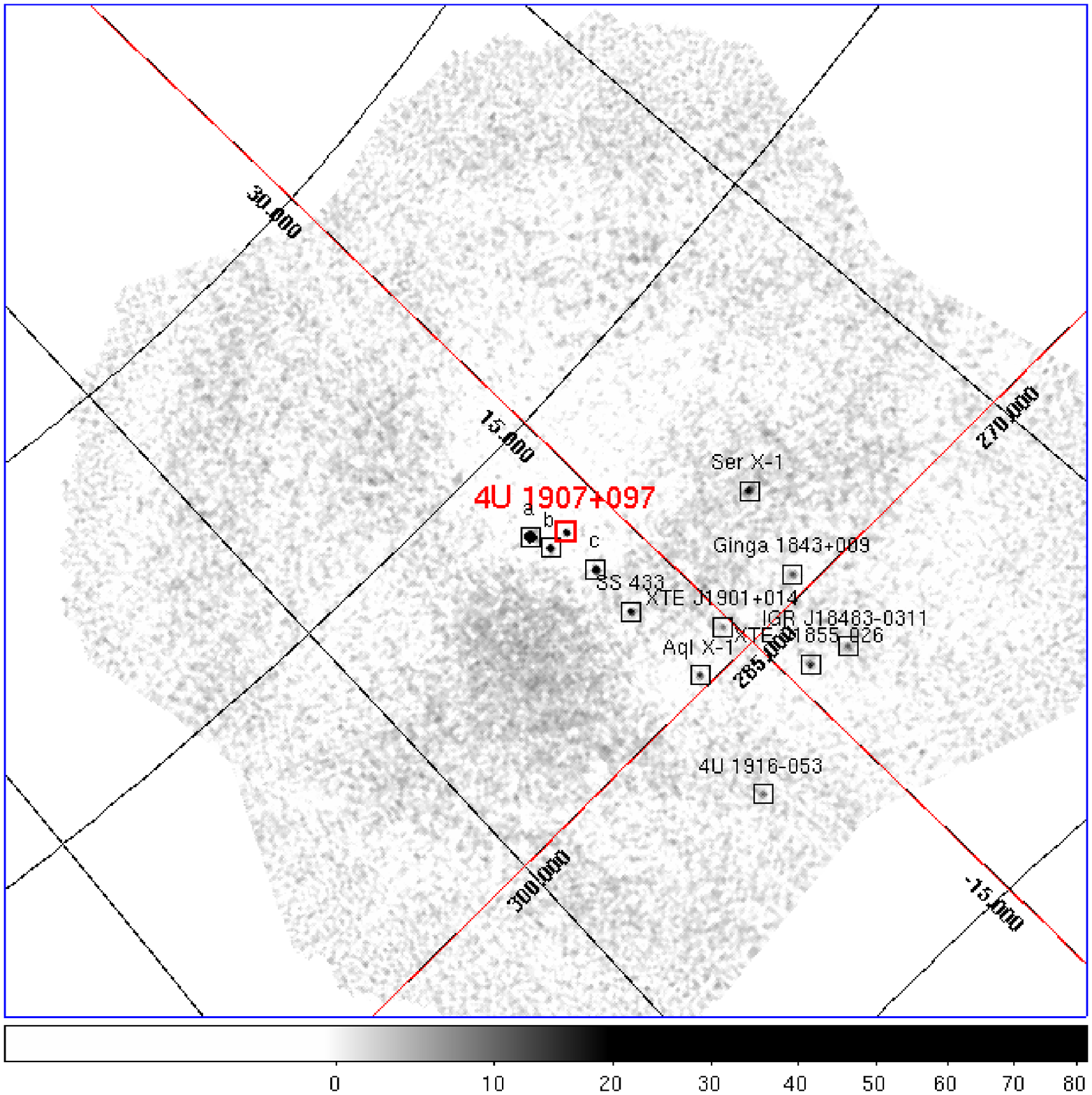}{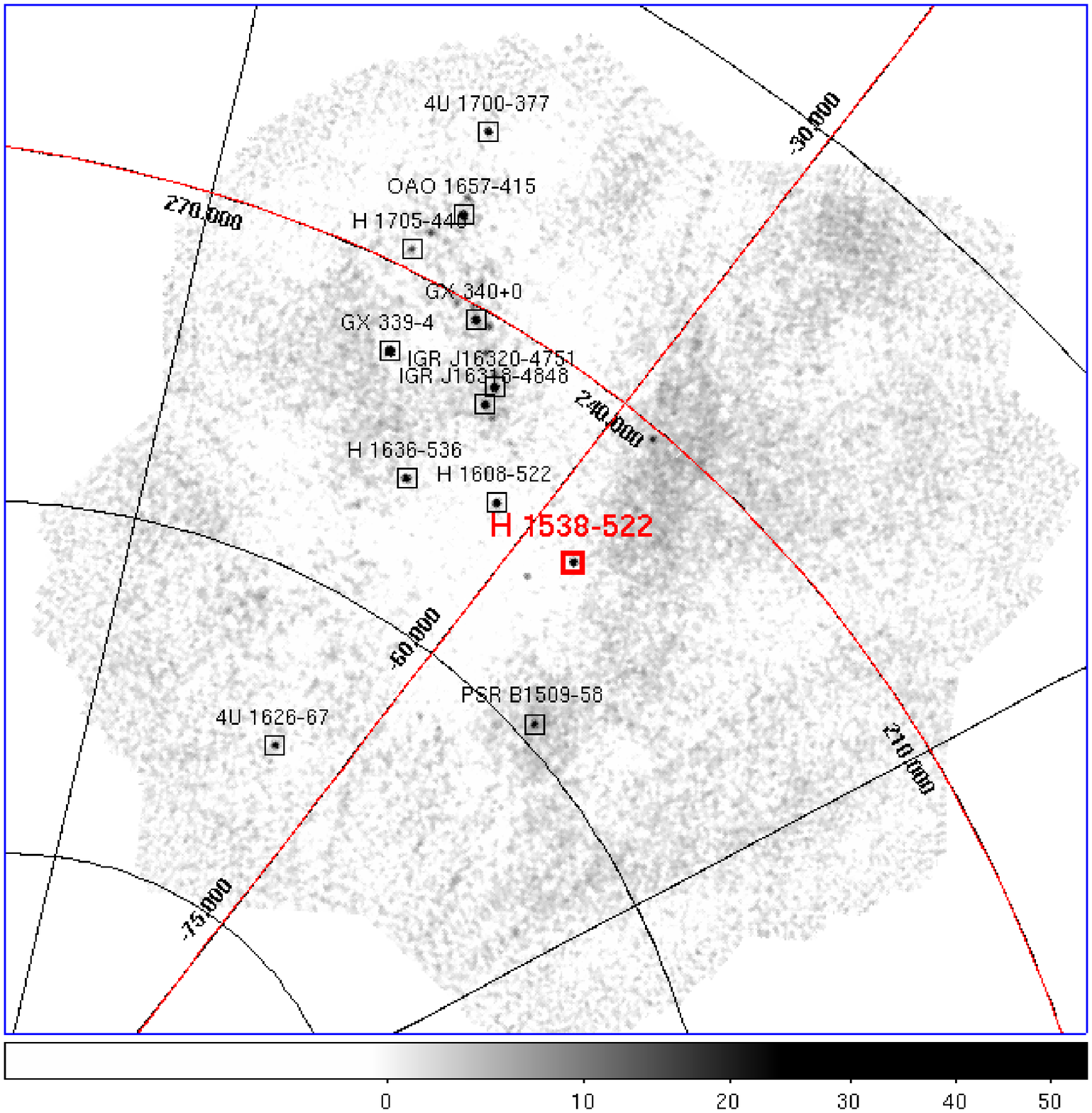}
  \caption{The combined 20-40\,keV significance mosaics of the sky around \nin (left)
    and \fif (right). Displayed coordinates are right ascension and declination
    in degrees. For clarity, the closest sources to \nin have been labeled with
    letters; they are (a) GRS 1915+105, (b) IGR J19140+0951, and (c) 4U 1909+07. While
    both images cover the full $27^\circ \times 27^\circ$ partially-coded field of view of
    the ISGRI telescope, the SCWs that were selected to form these images all have the
    sources within the $9^\circ \times 9^\circ$ \emph{fully}-coded field of view. Both
    sources are clearly detected in the combined mosaics.}
  \label{fig:double_mosaic}
\end{figure*}

Coded-mask detectors such as those used by ISGRI and JEM-X work by deconvolving
the ``shadowgram'' produced by opaque elements in the mask placed over the
detector. While this allows imaging to be performed in the normally difficult
hard X-ray band, it also means the illuminated pixels in the detector receive
light from every unocculted point in the field of view, and as such every
significant source in the FOV must be accounted for when determining the
background for a given single source. We followed the recommendation of
  the ISDC and provided a catalog of all sources with detection significance
  greater than $6.0\sigma$ in the single SCW's image. We provide the
  resulting list of sources in \ref{tab:incl}, along with their 20-40\,keV
counting rate in the combined mosaic of all SCWs. The 20-40 keV ISGRI mosaic
images from all SCWs are provided in Figure \ref{fig:double_mosaic}.

\begin{deluxetable}{crcr}
  \renewcommand{\arraystretch}{0.5}
  \tablecolumns{4}
  \tablewidth{0pt}
  \tablecaption{Sources included in the background subtraction routine for \fif
  and \nin, with 20$-$40\,keV ISGRI counting rates}
  \tablehead{ \colhead{Source} & \colhead{counts\ s$^{-1}$\tablenotemark{a}} & \colhead{Source} & \colhead{counts\ s$^{-1}$} }
  \startdata
    \fif     & $3.06$ & \nin              & $1.42$ \\
    \hline \\
    4U 1700$-$377     & $32.4$ & GRS 1915+105            & $44.9$\\
    GX 339$-$4        & $22.2$ & 4U 1909+07              & $2.26$\\
    OAO 1657$-$415    & $11.3$ & 4U 1916$-$053           & $1.76$\\
    4U 1626$-$67      & $6.70$ & XTE J1855$-$026         & $1.30$\\
    IGR J16320$-$4751 & $5.03$ & Ser X-1               & $1.24$\\
    GX 340+0          & $4.78$ & SS 433                  & $1.23$\\
    H 1608$-$522      & $4.25$ & IGR J19140+0951         & $1.12$\\
    H 1636$-$536      & $3.17$ & IGR J18483$-$0311       & $0.92$\\
    IGR J16318$-$4848 & $3.16$ & Aql X-1               & $0.79$\\
    H 1705$-$440      & $2.23$ & Ginga 1843+009          & $0.69$\\
    PSR B1509$-$58    & $1.42$ & XTE J1901+014           & $0.39$\\
    4U 1630$-$47      & ND\tablenotemark{b}     & SWIFT J1753.5$-$0127    & ND \\
    XTE J1652$-$453   & ND     & 4U 1812$-$12            & ND \\
    \nodata           & \nodata& GX 17+2                 & ND \\
    \nodata           & \nodata& PSR J1846$-$0258        & ND \\
    \nodata           & \nodata& 3A 1850$-$087           & ND 
  \enddata
  \tablenotetext{a}{20-40 keV ISGRI count rates}
  \tablenotetext{b}{ND: Not detected in mosaic image}
  \label{tab:incl}
\end{deluxetable}
\renewcommand{\arraystretch}{1.0}

\subsection{\fif}
\label{sec:H1538_obs}
Our ISGRI dataset comprises all publicly available data where \fif was within
its $4.5^{\circ}$ fully-coded field of view, extending from January 2003 through
April 2010. The JEM-X data are selected in a similar fashion, taking all windows
within the $2.4^{\circ}$-radius fully-coded field of view. There is a 3-year gap in the
\textit{INTEGRAL} coverage of \fif, with no observations between late 2005 and
late 2008. In the earlier data, the total ISGRI exposure is 715 ks, with 638 and
210 ks for JEM-X 1 and 2, respectively; the later dataset totals 417 ks of ISGRI
exposure, 411 ks of JEM-X 1, and 172 ks of JEM-X 2.

\subsection{\nin}
\label{sec:4U1907_obs}
We analyzed science windows beginning in November 2007 (\textit{INTEGRAL}
revolution 608) for \nin, overlapping slightly with the selection of
\citet{sahiner_comprehensive_2012}. Selecting SCWs according to the same
criteria as used for \fif amounts to 892 ks of ISGRI exposure, 908 ks of
JEM-X 1, and 507 ks of JEM-X 2. The average flux from \nin is considerably
lower than \fif, and is indeed at what appears to be close to a historic low
for the source (comparable to that seen by \citet{makishima_1907_1999} and
\citet{coburn_magnetic_2001}). As a result, it is difficult to provide
strong constraints on its spectral parameters.

\section{Timing Analysis}
\label{sec:timing}

We extracted 10\,s binned \textit{ISGRI} lightcurves in the 20-40\,keV energy
band with the OSA 10.0 analysis pipeline. The only departure from the standard
procedure was the use of the alternate \textit{ii\_light} tool for lightcurve
extraction, as the standard pipeline's lightcurve extraction routine is not
recommended for binning times less than 60\,s \citep{osa_ibis}. After
extraction, the lightcurves were barycentered using the \textit{barycent} tool
also provided in the OSA package.

Due to the low signal-to-noise and few pulsations covered by the individual
$\sim\!2$\,ks lightcurves, many lightcurves had to be appended to each other
before any period determination could be made. The composite lightcurve was
then corrected for the orbital motion of the source, using the best
available orbital parameters. As the pulse profile of the source is
non-sinusoidal and the composite lightcurve contains a moderate number of
gaps, epoch folding \citep{leahy_epfold_1983,larsson_epfold_1996} was used
for the period search, with 1$\sigma$ errors determined according to
\citet{larsson_epfold_1996}.

\subsection{\fif}
\label{sec:H1538_timing}
\begin{figure*}[ht]
  \begin{center}
    \plotone{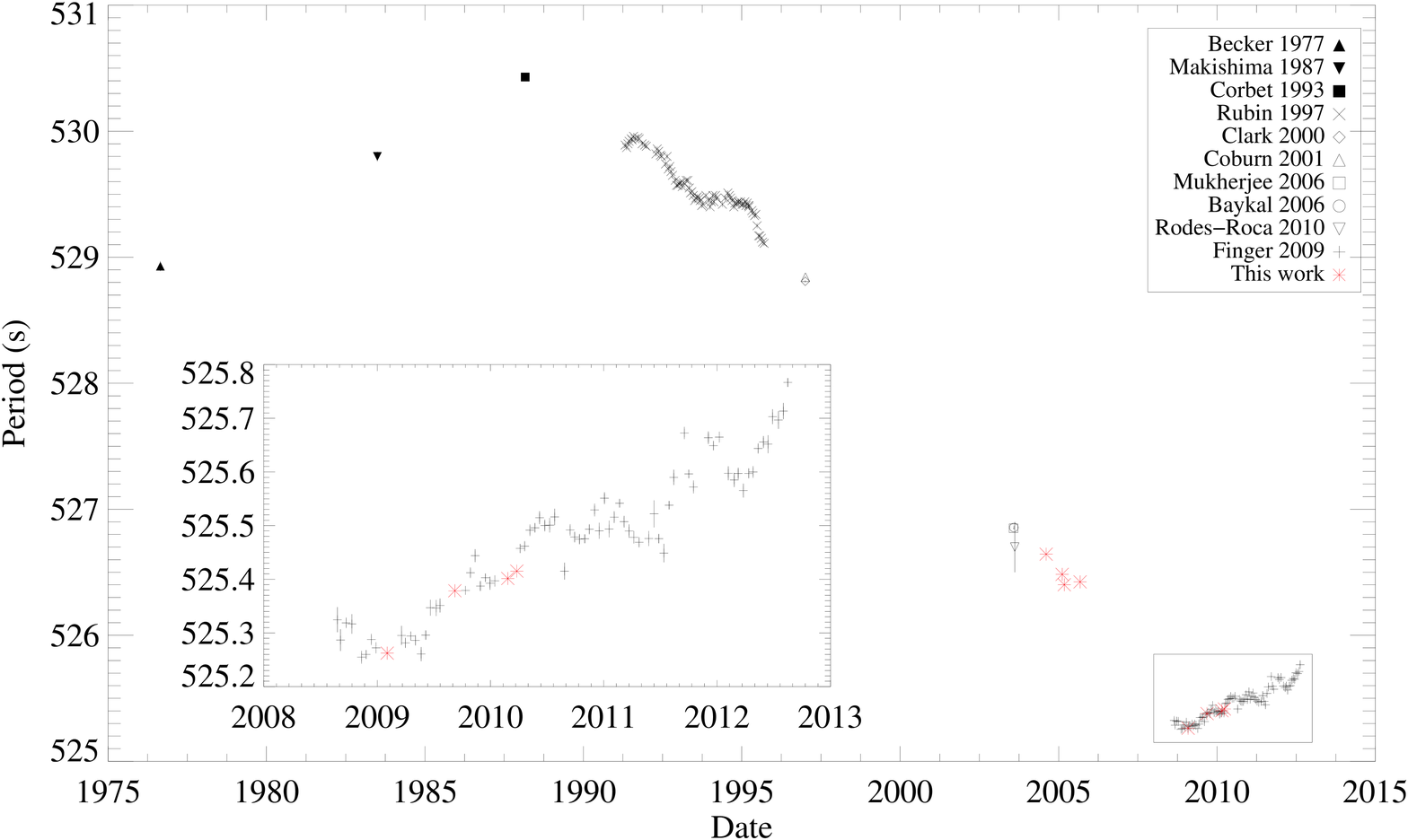}
  \end{center}
  \caption{Pulse period evolution of \fif. Binary-corrected periods found in
    this work are represented by red asterisks, with previous measurements
    \citep{becker_1538_1977,makishima_spectra_1987,corbet_orbit_1993,rubin_observation_1997,clark_orbit_2000,baykal_recent_2006}
    and recent \textit{Fermi} GBM data \citep{finger_gbm_2009} in black. Error
    bars are generally smaller than data points. {\fif} evidently underwent a
    torque reversal sometime in late 2008 or early 2009.  The final four
    datapoints are in very good agreement with the \textit{Fermi} GBM data;
    this agreement is detailed in the inset.}
  \label{fig:H1538_period}
\end{figure*}

For \fif, we used the elliptical orbital parameters provided by
\citet{mukherjee_orbital_2006}, although using \citet{clark_orbit_2000}'s
circular solution did not result in any significant change in the pulse period.
For convenience, these parameters are listed in Table \ref{tab:H1538_orb}.
Lightcurves were grouped by requiring that no gaps larger than 4 days be
present, in order to minimize the effect of large gaps in the data when
performing epoch folding. Eight sets of SCWs using this grouping have sufficient
time and statistics to return clear single peaks in the epoch folding results;
we summarize these results in Table \ref{tab:H1538_period}, and the source's
pulse period history is updated in Figure \ref{fig:H1538_period}. Pulse periods
from the earlier data, prior to 2008, follow roughly the same spin-up trend as
originally seen by \citet{rubin_observation_1997}, with the period dropping at a
rate of $\sim\!0.2$\,s\,yr$^{-1}$. However, our measured pulse periods post-2008
reveal a new spin-down trend of $\sim\!0.13$\,s\,yr$^{-1}$. This conclusion is
supported by the results of the \textit{Fermi} Gamma-ray Burst Monitor Pulsar
Project \citep{finger_gbm_2009}\footnote{Results for \fif can be found at \url{http://gammaray.nsstc.nasa.gov/gbm/science/pulsars}}, which we also include in
Figure \ref{fig:H1538_period}. The pulse shape in the 20-40\,keV band is
dominantly single-peaked; the disappearance of the secondary peak in the
$\gtrsim 20$\,keV pulse profile has been seen before by
\citet{clark_discovery_1990} and \citet{robba_bepposax_2001}, who suggested it
may be due to the presence of the $22$\,keV CRSF.  The pulse profiles from all
observations do not vary significantly from group to group, with the RMS values
of the difference between profiles $\sim$20 - 50 \% lower than the average noise
in the profiles. This persists across the torque reversal.

\begin{deluxetable}{llrr}
  \tablecolumns{4}
  \tablewidth{0pt}
  \tablecaption{Orbital parameters}
  \tablehead{\colhead{} & \colhead{} & \colhead{\fif\tablenotemark{a}} & \colhead{\nin\tablenotemark{b}}}
  \startdata
    $ a\sin(i) $           & lt-s     & $ 53.1 \pm 1.5 $ & $83 \pm 4$ \\
    $ e $                  & \nodata  & $ 0.18 \pm 0.01 $ & $0.28 ^{+0.10}_{-0.14}$ \\
    $ P_{\mathrm{orb}} $   & d        & $ 3.728382 \pm 0.000011 $ & $8.3753^{+0.0003}_{-0.0002}$ \\
    $ T_{\pi/2} $          & MJD      & $ 52851.33 \pm 0.01 $ & $50134.76^{+0.16}_{-0.20}$ \\
    $ \omega_{\mathrm{d}} $& \nodata  & $ 40^{\circ}  \pm 12^{\circ} $ & $330 \pm 20^{\circ}$
  \enddata
  \tablenotetext{a}{\citet{mukherjee_orbital_2006}}
  \tablenotetext{b}{\citet{intzand_1907_1998}}
  \label{tab:H1538_orb}
\end{deluxetable}

\begin{deluxetable}{cr}
  \tablecolumns{2}
  \tablewidth{0pt}
  \tablecaption{Periods for \fif determined via epoch folding.}
  \tablehead{\colhead{MJD} & \colhead{Period (s)}}
  \startdata
    $53222.176-53230.047$ & $ 526.644 \pm  0.003 $ \\
    $53407.638-53415.310$ & $ 526.483 \pm  0.003 $ \\
    $53429.742-53439.492$ & $ 526.401 \pm  0.002 $ \\
    $53616.904-53620.899$ & $ 526.424 \pm  0.007 $ \\
    $54855.844-54870.322$ & $ 525.262 \pm  0.002 $ \\
    $55073.587-55090.517$ & $ 525.379 \pm  0.001 $ \\
    $55250.376-55258.045$ & $ 525.401 \pm  0.004 $ \\
    $55264.806-55288.806$ & $ 525.415 \pm  0.001 $
  \enddata
  \label{tab:H1538_period}
\end{deluxetable}

\begin{figure}
  \begin{center}
    \plotone{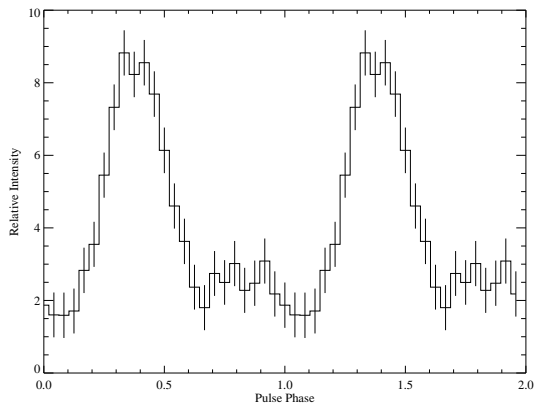}
  \end{center}
  \caption{A typical 20-40\,keV pulse profile for \fif. The average pulse
    shape is strongly single-peaked in this energy band. This profile is for
    the first epoch listed in Table \ref{tab:H1538_period}; the variations
    between pulse profiles for the different epochs listed are not
    statistically significant.}
  \label{fig:H1538_pulse}
\end{figure}

\subsection{\nin}
\label{sec:4U1907_timing}
We began our \nin timing analysis with data from \textit{INTEGRAL} revolution
608, roughly where the \textit{INTEGRAL} analysis of
\citet{sahiner_comprehensive_2012} ended. The data were binary corrected
according to the parameters determined by \citet{intzand_1907_1998}, listed in
Table \ref{tab:H1538_orb}. During the time encompassed by our analysis, the
source was considerably fainter when compared to previous studies using
\textit{INTEGRAL} \citep{fritz_1907_2006,sahiner_comprehensive_2012}, and
was significantly ($\sim\!40\%$) dimmer than \fif  (compare the 5-100\,keV
fluxes for \fif in Tables \ref{tab:H1538_date} and \ref{tab:H1538_lum} to
the equivalent flux for \nin in Table \ref{tab:4U1907_full_spectrum}). When
epoch folding, no selection of SCWs from the new \textit{INTEGRAL} data
produce $\chi^{2}$ distributions that are comparable to our results for
\fif, and all error-estimation procedures return errors of $\sim\!1$\,s,
compared to $\sim\!0.001$\,s for \fif. As a consistency check, SCWs from
\textit{INTEGRAL} revolutions 608 through 623, already reported by
\citet{sahiner_comprehensive_2012}, were analyzed in the same manner; for
this data subset alone, the epoch folding returns a statistically
significant result that is consistent with
\citeauthor{sahiner_comprehensive_2012}'s. We thus do not report any
\textit{INTEGRAL}-derived pulse periods for \nin due to the faintness of the
source.

\section{Spectral Analysis}
\label{sec:spectral}
We extracted ISGRI and JEM-X spectra for \fif and \nin on a SCW-by-SCW basis
using the OSA 10.0 software, combining the individual SCW spectra with the
\texttt{spe\_pick} tool provided in the OSA package. The extracted spectra were
then modeled in \textit{XSPEC} version 12.7.1 \citep{arnaud_xspec_1996}. All
error bars are for the 90\% single-parameter confidence interval unless
otherwise indicated, and 5-100\,keV model fluxes with 90\% error bars were
computed using \textit{XSPEC}. The spectra for both sources descend into noise
above $\sim$50-80\,keV, depending on the particular subset of data being
analyzed. Spectra were regrouped to provide roughly similar error bars
for data above the noise threshold.

We present spectral fits for two continuum models. The PLCUT model (in
\emph{XSPEC}, POWERLAW*HIGHECUT) is a piecewise function, with a power-law
multiplied by a high energy exponential cutoff above some cutoff energy
$E_{\mathrm{cut}}$: 

\begin{equation} 
  PLCUT(E) = \left\{ \begin{array}{cc} 
  A \times E^{-\Gamma} & E < E_{\mathrm{cut}} \\
  A \times E^{-\Gamma}\exp\left(\frac{E_{\mathrm{cut}} - E}{E_{\mathrm{fold}}} \right) & E \ge E_{\mathrm{cut}} \\
  \end{array} \right.
  \label{eqn:plcut} 
\end{equation}
where $A$ is the model flux in erg cm$^{-2}$ s$^{-1}$ at 1 keV and
$E_{\mathrm{fold}}$ is the $e$-folding energy that turns on at energies higher
than the cutoff energy $E_{\mathrm{cut}}$. The CUTOFFPL continuum is PLCUT with
the $E_{\mathrm{cut}}$ parameter frozen at zero. Since this means the CUTOFFPL
spectrum is being exponentially attenuated at all energies, it generally has a
significantly lower photon index $\Gamma$, but its continuous nature means it
lacks the sharp corner that in PLCUT can produce spurious residuals around
the cutoff energy. However, CUTOFFPL on its own results in high $\chi^{2}$
values for spectral fits in some cases, requiring an additional broad Gaussian
component to be added to the spectral model. A feature like
this at $\sim\!10$\,keV has been seen before in both \fif
\citep{coburn_magnetic_2001} and \nin
\citep{mihara_thesis_1995,coburn_magnetic_2001,rivers_comprehensive_2010} as an
emission feature, as well as many other HMXBs both with and without cyclotron
features \citep[although it occasionally can be modeled as an absorption feature, see
e.g., ][]{muller_sleeping_J1946_2012}, and is likely due to the simplistic
nature of our continuum models in light of the complex physics of the actual
accretion column.

There are two other commonly used continuum models, the Fermi-Dirac cutoff
model FDCUT and the two-power-law NPEX, but neither model was capable of
producing a stable fit in either source, generally returning many
unconstrained parameters. We did not incorporate any photoelectric absorption,
as it is not strongly pronounced at energies above a few keV, nor did we
include the Fe K line at 6.4\,keV \citep{rodes-roca_detecting_2010}, as the
energy binning needed to obtain usable statistics is too coarse to observe the
line.

Both sources exhibit absorption-like features in the $\sim\!20$ and
$\sim\!40-50$\,keV range. These are typically identified as CRSFs, and a brief
description of the mechanism for their production is presented in \S
\ref{sec:discussion}. They typically appear as broad ($\sigma \gtrsim 3$\,keV),
relatively shallow absorption-like features, usually modeled by multiplicative
Gaussians or pseudo-Lorentzians with negative intensity. We use a local
\textit{XSPEC} model, GAUABS, a Gaussian optical depth profile defined by

\begin{eqnarray}
  \mathrm{GAUABS}(\tau) &=& F_{0} e^{-\tau\left(E\right)} \\
  \tau(E) &=& \tau_{0}\exp\left( - \frac{\left(E-E_{0}\right)^{2}}{2\sigma^{2}}\right)
  \label{eqn:gauabs}
\end{eqnarray}

The broad width and shallowness of these lines makes their detection and
constraint difficult in the exponentially-dropping spectra of HMXBs, especially
when it comes to the higher harmonics which lie far above the cutoff energy of
the spectrum. In particular, the width of the harmonic was unconstrained in some
cases, and we opted to fix it at its fitted value when determining errors. In
very low-flux datasets, the harmonic CRSF is often not detected.

\subsection{\fif}
\label{sec:H1538_spectral}

\subsubsection{Date-selected spectra}
\label{ssub:H1538_date}

The three-year gap in \textit{INTEGRAL} coverage of \fif, along with the
presence of the torque reversal of late 2008/early 2009, presents an obvious
point at which to split the dataset, so we analyze the 2003--2005 and
2008--2010 spectra separately.

\begin{figure*}[t]
  \includegraphics[width=0.5\textwidth]{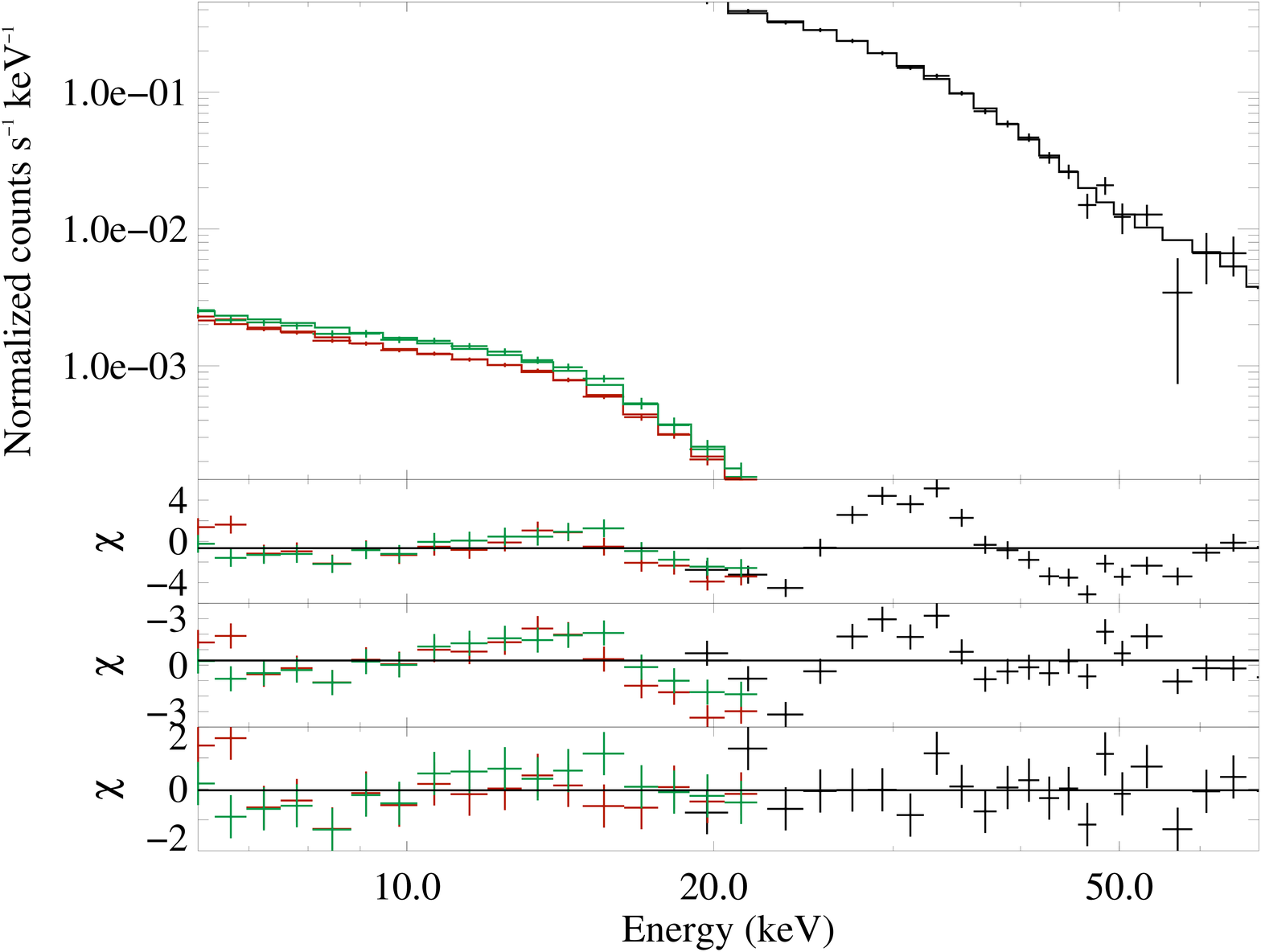}
  \includegraphics[width=0.5\textwidth]{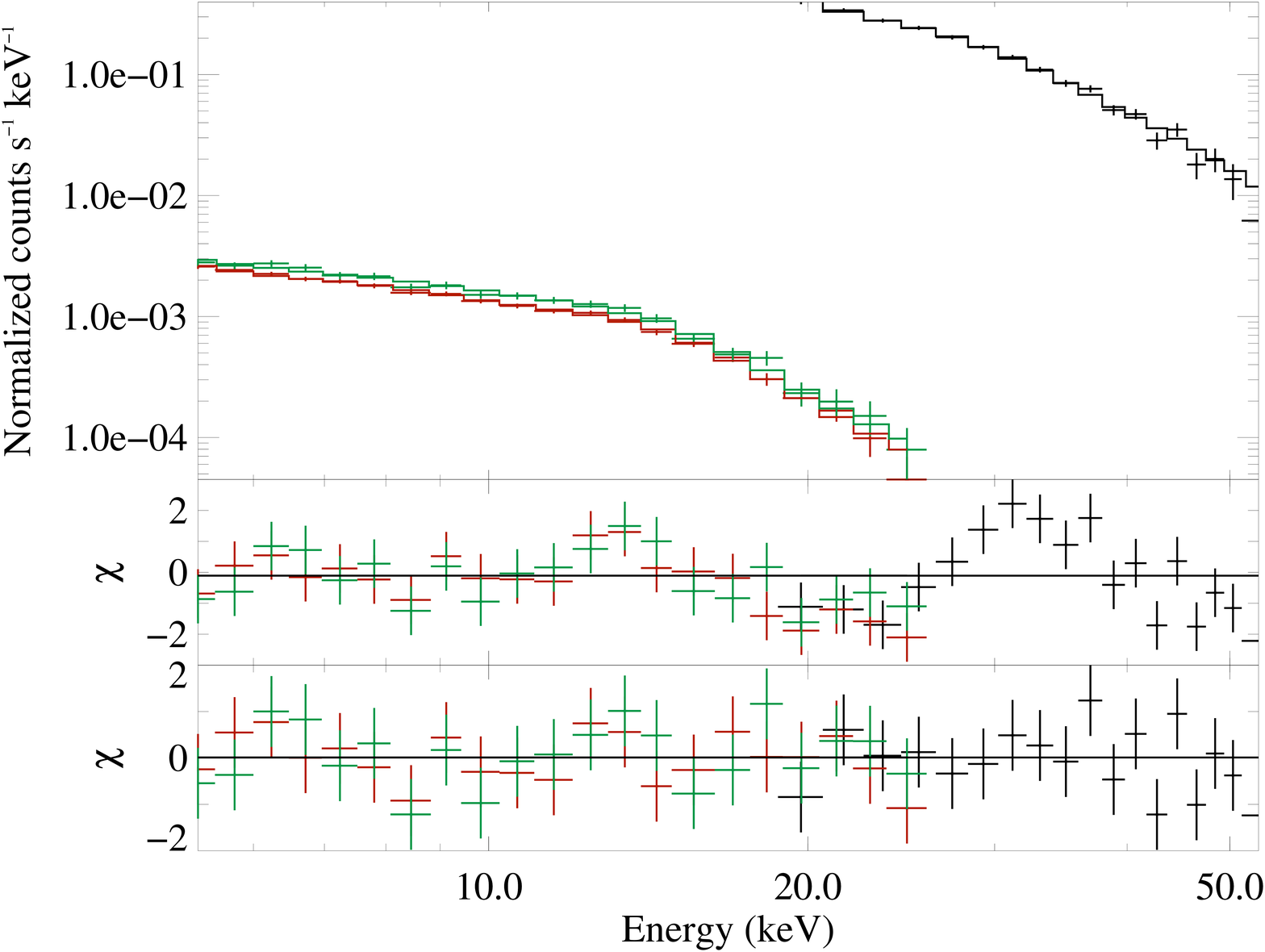}
  \caption{Date-selected counts spectra and models for \fif using the PLCUT
    continuum. Black points indicate the ISGRI spectrum; red and green are
    JEM-X 1 and 2, respectively. The left panel displays the earlier
    dataset, comprising data from 2003--2008, with residuals for the PLCUT
    continuum with, from top to bottom, no CRSFs, one CRSF at $\sim\!50$\,keV,
    and CRSFs at $21.5$\,keV and $50$\,keV. The right panel contains the later
    dataset, spanning the years 2008--2010, and the residuals are for the
    continuum with no CRSFs (top) and one CRSF at $21.1$\,keV (bottom).}
  \label{fig:H1538_date}
\end{figure*}

Each dataset was fit with the two aforementioned continuum models.
Each continuum was modified by one or two Gaussian-profile absorption features
and a constant multiplier, which was frozen at unity for ISGRI spectra and
allowed to vary for the JEM-X spectra in order to account for calibration
differences between instruments. The two continuum models provide similar
$\chi^{2}$ statistics, with CUTOFFPL+BUMP being slightly superior in most
cases. However, when calculating errors, it was necessary to fix the width of
the 10-keV feature in the later dataset at the value found in the
earlier dataset, as the decreasing sensitivity of the JEM-X telescopes at
higher energies combined with the presence of the CRSF at $\sim\!22$\,keV
resulted in unconstrained fit parameters. We will refer primarily to the PLCUT
continuum results in our discussion due to this limitation. Neither continuum
model shows any significant variation across the torque reversal. We list the
parameters for the spectral fits for both continua in Table
\ref{tab:H1538_date}, and the spectra for the PLCUT continuum are plotted in
Figure \ref{fig:H1538_date}.

We clearly detect the fundamental CRSF at $21.5^{+0.5}_{-0.6}$\,keV in the early
dataset and $21.1 \pm 0.8$\,keV in the late. Adding the feature in the
early spectra decreases the $\chi^{2}/\mathrm{DOF}$ statistic from $327/50$ to
$70/47$, while in the late dataset, $\chi^{2}/\mathrm{DOF}$ drops from
$84/49$ to $33/46$. There are no statistically significant differences in
the fundamental CRSF across the torque reversal. The harmonic CRSF is much less
significantly detected in the early dataset: its addition at $50^{+5}_{-4}$\,keV
lowers the $\chi^{2}/\mathrm{DOF}$ to $52/44$. In the later data, the feature is
not detected; adding a $\sim\!50$\,keV GAUABS component reduces the $\chi^{2}$
by a negligible amount, and generally has unconstrained width and depth, so we
leave the feature out of our model for the late data.  This non-detection is
likely a product of the fact that we are adding together all spectra from after
the torque reversal, and the lower-luminosity observations are driving down the
signal-to-noise ratio. This can be more clearly seen in the following section.

\input{H1538_table_date_rebinned}

\subsubsection{Luminosity-selected spectra}
\label{ssub:H1538_lum}

To more closely examine the differences between the earlier and later spectra
for \fif, we further split the date-selected SCWs into high, middle, and
low-luminosity bins by taking a 1$\sigma$ region around the average-brightness
SCW to be the middle-luminosity bin. The 5$-$100\,keV flux of the source varies
by a factor of $\sim\!3$ from the lowest- to highest-luminosity bin. The
best-fit parameters for these datasets for two continuum models are presented
in Table \ref{tab:H1538_lum}. Both continuum models again provide roughly
similar $\chi^{2}$ statistics for each dataset. The earlier mid-luminosity bin
was the only dataset in the luminosity-selected data that showed a clear need
for a 10-keV feature; however, it was necessary to freeze the width of the
feature at its fitted value to obtain a stable fit, due to the
interaction between the $\sim 22$\,keV CRSF and the ``bump.'' We additionally
froze the width of the harmonic CRSF in all but one model for the
mid-luminosity datasets, as a dearth of counts at high energy often led this
parameter to be unconstrained. In these cases, we froze the value of
the width to the value found in the same model for the corresponding
high-luminosity dataset, the exception being the later mid-luminosity spectrum,
where the 16\,keV width of the later high-luminosity data did not produce a
good fit. In that case, the width was frozen to the value found in the earlier
high-luminosity data. The harmonic CRSF was not detected in the low-luminosity
dataset.

In the PLCUT continuum, E$_{\mathrm{fold}}$ is the only parameter showing a
significant correlation with luminosity, with a slope of $1.2 \pm
  0.8$\,keV/($10^{36}$\,erg\,s$^{-1}$) (90\% confidence). The correlation is
significant at the p = 0.02 level. The other spectral parameters generally do
not show clearly significant trends; $\Gamma$ and $E_{\mathrm{cut}}$
have slight correlations, but only at the $1\sigma$ level. The missing
harmonic CRSF is consistent with a lack of statistics at energies above
$\sim\!40$\,keV in the low-luminosity datasets; attempting to fit a line with
energy and width frozen at the value from the mid-luminosity bin results in an
unbounded depth, and so we do not include the harmonic CRSF in the model for
the low-luminosity data.

\input{H1538_table_luminosity_rebinned}

\subsection{\nin}
\label{sec:4U1907_spectral}

\begin{figure}[h]
  \begin{center}
    \plotone{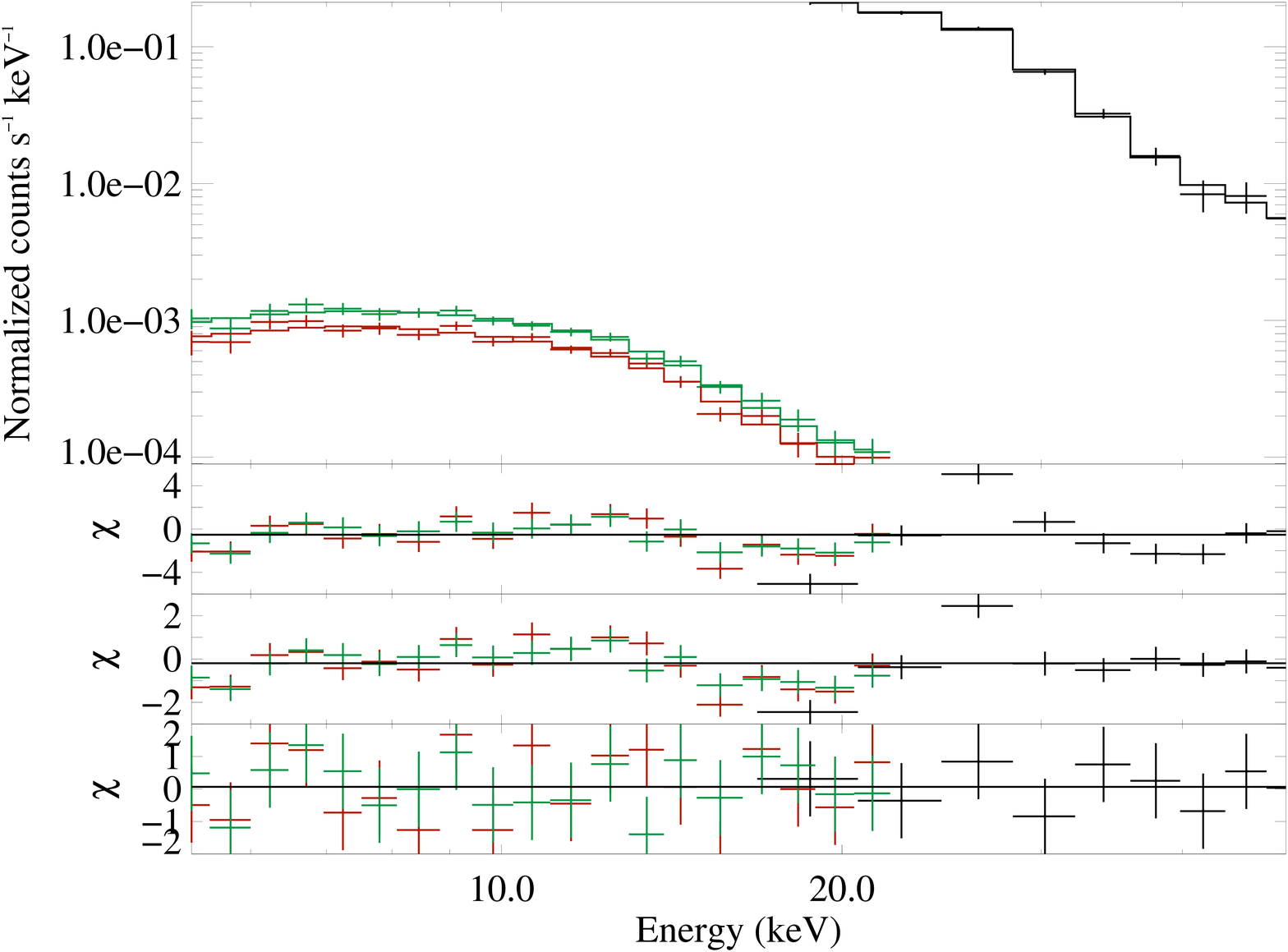}
  \end{center}
  \caption{The spectrum of \nin with the best-fitting model, a PLCUT continuum
    modified by two GAUABS absorption features at $18.5$ and $38$\,keV. The
    second panel displays the residuals from a pure PLCUT model, the third after
    only the harmonic CRSF has been added, and the lower panel displays the
    residuals with both CRSFs.}
  \label{fig:4U1907_full_spectrum}
\end{figure}

The SCWs for \nin are overall more evenly spread throughout
\textit{INTEGRAL}'s mission, with no significant changes in pulse period
derivative, so there is no logical point in time to define a date-based
separation as we have done for \fif. The best-fit spectral parameters for
all analyzed \nin SCWs are presented in Table
\ref{tab:4U1907_full_spectrum}, with the folded spectrum and model residuals
for the PLCUT continuum plotted in Figure \ref{fig:4U1907_full_spectrum}.
Previous analyses
\citep{fritz_1907_2006,rivers_comprehensive_2010,sahiner_comprehensive_2012}
have been performed using the PLCUT, FDCUT, and NPEX continua, but as
is stated in \S \ref{sec:spectral}, we could not obtain useful fits for the
latter two.  There was no evidence in the residuals for a bump feature at
$\sim\!13$\,keV as has been seen in spectra from e.g., \textit{Suzaku}
\citep{rivers_comprehensive_2010}; an attempt at adding in the feature
produced minimal ($\ll 1$) improvements in $\chi^{2}$ along with very poorly
constrained energy, width, and height. Without the feature, the CUTOFFPL
continuum produces equally acceptable results as compared to the PLCUT
continuum.

The addition of the fundamental CRSF at $\sim\!18$\,keV improves the fit
significantly, lowering the $\chi^{2}$ by $\sim\!100$. This feature can be seen
primarily in the JEM-X residuals, while ISGRI, with its 20\,keV lower bound, can
only resolve its upper edge. The harmonic is much less significantly detected at
$\sim\!38$\,keV, producing a drop of $\sim\!10$\ in $\chi^{2}$ in both models.
Adding the harmonic \textit{sans} the fundamental produces a drop of $\sim\!20$.
While the spectrum is already well-fit with only one CRSF, we include both in our
final models on account of their well-established existence in the literature
\citep{cusumano_1907_1998,makishima_1907_1999,fritz_1907_2006,rivers_comprehensive_2010},
and the fact that we are still capable of constraining the harmonic CRSF. Our
measurement of the fundamental CRSF is compatible with the measurements by
\citet{makishima_1907_1999} and \citet{coburn_magnetic_2001}, which were made
using observations of the source at a similar flux, although, as will be
discussed in \S \ref{sec:discussion}, \citeauthor{makishima_1907_1999}'s
measurement was made using a different model for the line, which could bring
that result out of compatibility with ours.

An attempt was made to split \nin's spectra into luminosity bins as we did for
\fif, but the poorer statistics for \nin made this unfeasible, with satisfactory
fits only obtained for the brightest luminosity bin. This difficulty is
illustrated by the considerably lower 5-100\,keV flux of \nin compared to \fif -
the overall flux from the full \nin dataset was $\sim\!40\%$ lower than the flux
for \fif, and was in fact considerably lower than the flux from our
lowest-luminosity dataset for \fif (compare Table \ref{tab:4U1907_full_spectrum}
with Tables \ref{tab:H1538_date} and \ref{tab:H1538_lum}).

\input{4u1907_table_rebinned}

\section{Discussion}
\label{sec:discussion}

\subsection{Torque Reversal}
\label{ssec:discuss_torque}
Our pulse period measurements for \fif show evidence for a torque reversal
in late 2008 or early 2009, although no significant spectral changes are
detected that could shed light on a mechanism for this change. Pulse period
evolution in accreting neutron stars like \fif is primarily driven by the
interaction between the accreting material and the magnetic field of the
star. As the field lines are locked into the neutron star, they rotate with
the star, and the infalling material can exert a torque on the star. Changes
in the accretion rate can thus produce changes in the pulse period, but the
relationship between accretion rate and torque can be very complex due to
the interaction between the accreting material, which may or may not form an
accretion disk, and the magnetic field of the neutron star. \fif is
primarily a wind-accretor, as the main-sequence companion \qvnor is likely
not overfilling its Roche lobe \citep{reynolds_optical_1538_1992}, so it is
unlikely that the accreted material comes in with sufficient angular
momentum to form a persistent disk, although there may be transient disk
formation. However, to our knowledge, no searches for features that might
indicate the existence of a transient disk (\textit{i.e.} a $\sim$few mHz
QPO) have been performed.

The wind-accreting nature of the source, while it can explain
short-timescale variations in pulse period, makes it more difficult to explain
the long-term spin-up or spin-down trends observed. The torque reversal in
particular is interesting: torque reversals observed in other sources have
typically included large shifts in luminosity and/or dramatic spectral shifts
\citep[see, e.g., 4U~1626$-$67, ][]{cameroarranz_torque1626_2010}. In \fif, we see
no shift in the X-ray luminosity across the reversal, and the spectral
parameters generally do not exhibit significant changes. Similar to 4U~1626$-$67
\citep{cameroarranz_torque1626_2010} and \nin \citep{inam_torque_2009}, the
spin-up and spin-down rates on either side of the reversal are similar: $1.4
\times 10^{-14}$\,Hz\,s$^{-1}$ for the current spin-down trend, compared to a
spin-up of $1.8 \times 10^{-14}$\,Hz\,s$^{-1}$ between 1990 and 2009. In the
case of \nin's most recent torque reversal, \citeauthor{inam_torque_2009}
suggested that the model of \citet{perna_spin_2006} may be useful, as they are
able to produce torque reversals without large changes in luminosity or
$|\dot{\nu}|$. However, this generally requires an Alfv\'{e}n radius comparable
to the corotation radius of the neutron star to produce the localized propellor
effect on which the \citeauthor{perna_spin_2006} model relies. \fif's slow
rotation speed means its $\sim\!10^{8}$\,m corotation radius is 1-2 orders of
magnitude larger than its Alfv\'{e}n radius, which is $\sim\!4 \times 10^{6}$\,m
if spherical accretion is assumed \citep{lamb_accretion_1973}, so the recycling
mechanism of the \citeauthor{perna_spin_2006} model is unlikely to be in effect.
Meanwhile, \citet{rubin_observation_1997} modeled \fif's spin-up trend of $1991 -
1995$ as simply the consequence of a random-walk in the pulse frequency
derivative (stemming from a random-walk in the accretion torque on the neutron
star), and proposed that this alone could be the source of the 1990 reversal,
which would allow for the constancy of the system across the reversals. However,
the nearly twenty year spin-up trend clearly indicates that the system is not
solely random-walking -- while for \citeauthor{rubin_observation_1997}, the
expectation value for the frequency shift was within a factor of three of the
observed shift, the most recent pulse period measurements are a factor of
$\sim\!8$ larger than the expected RMS excursion. Without observations
tracking the actual torque reversals (as was the case with \nin), it is
difficult to address the torque reversal in any more depth.

\begin{figure*}[ht!]
  \begin{center}
    \plotone{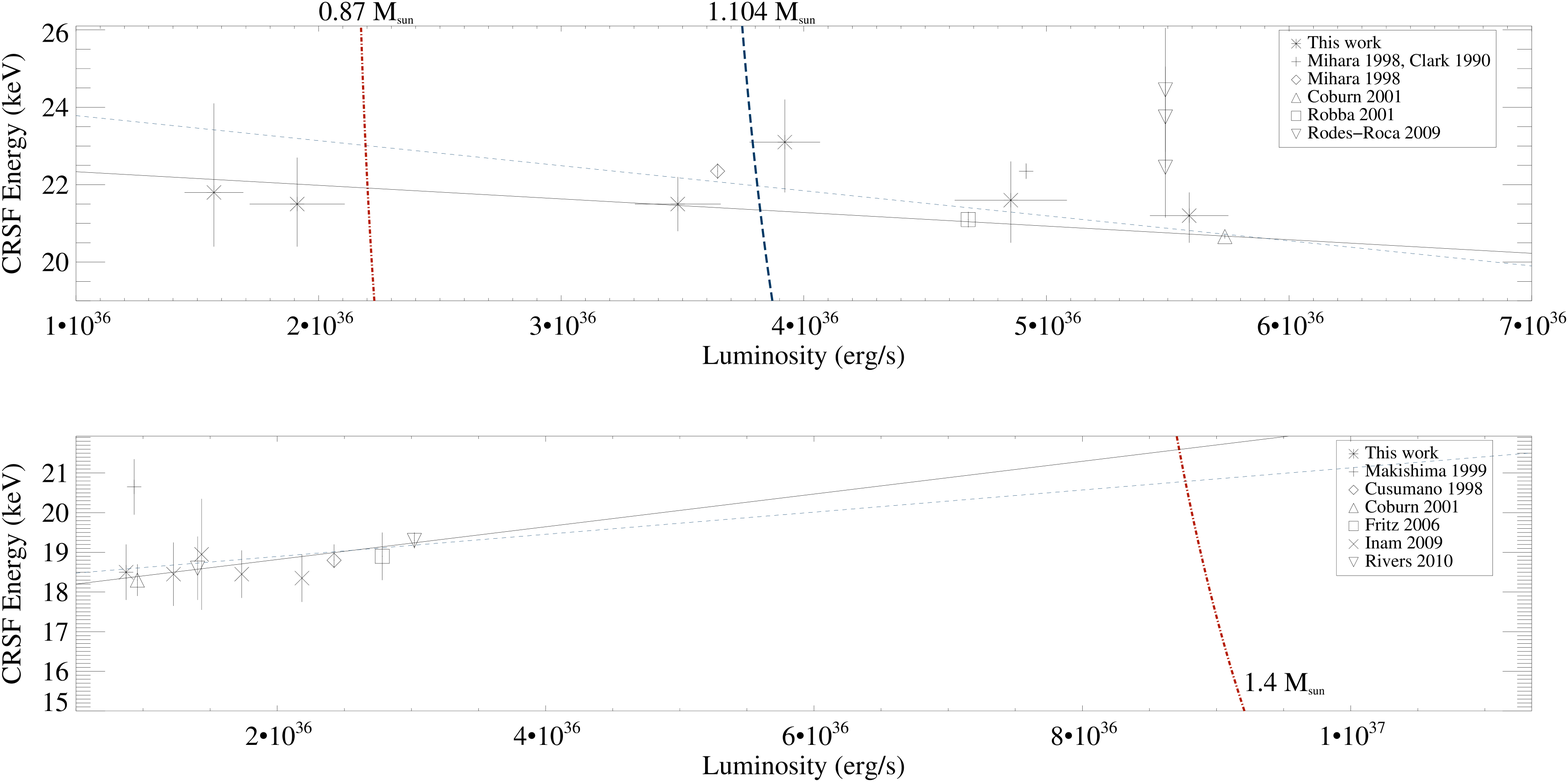}
  \end{center}
  \caption{Cyclotron line energy \textit{versus} luminosity for \fif (top
  panel) and \nin (bottom). Measurements from this work are represented
    by asterisks; the values and 90\% error bars are taken from models using
  the PLCUT continuum. Overplotted in heavy dashed and dot-dashed lines are
    curves for $L_{\mathrm{coul}}$ from \citet{becker_spectral_2012}; in the
    top panel, the red heavy dot-dashed line represents $L_{\mathrm{coul}}$ for
    $M_{X} = 0.87$\,\msol, while the blue dashed line uses $M_{X} =
    1.104$\,\msol.  Luminosities and values for $E_{cyc}$ from other works are
    included, along with 90\% confidence intervals. Linear fits to the data are
    also overplotted; due to the systematic differences between Gaussian- and
    Lorentzian-shaped models for the cyclotron lines, there are two curves for
    each source: the solid includes only sources that used Gaussian-profile
    models for the CRSFs, while the dotted line includes Lorentzian-modeled
    lines (those of \citet{mihara_cyclovar_1998}, \citet{clark_discovery_1990},
    \citet{rodes-roca_first_2009}, \citet{makishima_1907_1999}, and
    \citet{inam_torque_2009}) shifted upwards by 1.75\,keV.}
  \label{fig:l_vs_e}
\end{figure*}

\subsection{Cyclotron Line Variability}
\label{ssec:discuss_crsf}
We detect absorption-like features, modeled as lines with Gaussian optical
depth profiles, at $\sim\!22$ and $\sim\!50$\,keV in \fif and $\sim\!18$ and
$\sim\!38$\,keV in \nin, which we identify as CRSFs. These features are
produced by the quantization of cyclotron motion in the strong magnetic
field of the neutron star - photons with the appropriate energy and momentum
to excite an electron into a new Landau level will see an increased
scattering cross-section. This manifests as a set of absorption-like
features produced when photons are scattered out of the line of sight. For
magnetic fields of strength $\lesssim 4\times 10^{13}$\,G, the magnetic
field strength can be approximated from the energy of the fundamental
cyclotron line via the ``12-B-12'' rule: $E_{\mathrm{cyc}} =
(11.57\,\mathrm{keV})/(1+z) \times B_{12}$, where $z$ is the gravitational
redshift in the scattering region and $B_{12}$ is the magnetic field
strength in units of $10^{12}$\,G. If we assume our measured CRSF energies
reflect the magnetic field at the stellar surface, this implies a magnetic field
strength of $\sim 2.1-2.3 \times 10^{12}$\,G for \fif and $2.0 \times
10^{12}$\,G for \nin. Here, we have used masses for \fif of $0.87 \pm 0.07$ and $1.104
\pm 0.177$\,\msol as determined by \citet{rawls_mass_2011}, and we have assumed
a mass of $1.4$\,\msol for \nin. We have assumed a radius of $10$\,km for both
sources.

A more in-depth investigtion of the cyclotron line scattering region has been
carried out by \citet{becker_spectral_2012}, who predict the behavior of the
CRSF in response to the luminosity of the source. In the supercritical regime,
above the effective Eddington luminosity of the accretion column, they
demonstrate that the height of the scattering region should scale with
luminosity, producing a negative correlation between CRSF energy and luminosity,
as the scattering region moves up into regions of weaker magnetic field.
Meanwhile, in subcritical sources, the infalling material comes to a halt via a
combination of gas and radiation pressure. An increase in luminosity means less
gas pressure is necessary, lowering the scattering region and producing a
positive $E_{\mathrm{cyc}}$-luminosity correlation. At even lower luminosity,
below the Coulomb stopping limit $L_{\mathrm{coul}}$ defined by
\citeauthor{becker_spectral_2012}, the sum total gas and radiation pressure does
not halt the infalling material before it impacts the star, and the scattering
should happen at or near the stellar surface. Here, the predicted behavior of
the cyclotron line energy is somewhat uncertain, although
\citeauthor{becker_spectral_2012} suggest that there should not be any large
correlation of $E_{\mathrm{cyc}}$ with luminosity. 

Adopting distances of $6.4$ kpc for \fif \citep{reynolds_optical_1538_1992} and
$5$ kpc for \nin \citep{cox_1907_2005} and determining the luminosity based
on the 5-100\,keV flux, we calculated $L_{\mathrm{crit}}$ and
$L_{\mathrm{coul}}$ using \citeauthor{becker_spectral_2012}'s formulae and
found that both sources lie firmly in the subcritical ($L <
L_{\mathrm{crit}}$) regime, around or below $L_{\mathrm{coul}}$. The exact
location of \fif relative to $L_{\mathrm{coul}}$ varies depending on which
mass is used: with the very low-mass elliptical result, the measured
luminosities for the source lie mostly above the $L_{\mathrm{coul}}$ cutoff,
while the higher-mass circular solution has the source straddling the
$L_{\mathrm{coul}}$ line. With no published mass measurements for \nin, we
assume the canonical neutron star parameters of $M_{\mathrm{X}} =
1.4$\,\msol and $R_{\mathrm{X}} = 10$\,km, which places the source far below
the $L_{\mathrm{coul}}$ cutoff. While the value for $L_{\mathrm{coul}}$ can
vary widely with mass ($L_{\mathrm{coul}} \propto M_{\mathrm{X}}^{11/8}$),
unless \nin has a very low mass similar to \fif it is unlikely for it to
move above $L_{\mathrm{coul}}$ with an updated mass value. Our values for
CRSF \emph{vs.}\ luminosity are plotted against $L_{\mathrm{coul}}$ for both
sources in Figure \ref{fig:l_vs_e}. To supplement this work's results, we also
include the historical measurements of the energy of the primary CRSF for each
source, limited to cases where the luminosity or flux was provided or could be
derived from models. Based on our measured luminosity errors, we have
assumed an error of 3\% for these luminosity values.

Of course, when comparing results from different references, the issue of
model choice must be addressed. Several analyses of \fif and \nin used a
combination of the NPEX continuum and pseudo-Lorentzian line profiles for the
CRSFs. This choice of models results in overall lower measured energies for the
cyclotron lines and larger line widths compared to work that used Gaussians or
the GAUABS model we have used. A formal treatment of this difference would be
beyond the scope of this work: \citet{mihara_thesis_1995} provides the
analytical expression for the energy of the maximum of the pseudo-Lorentzian
model, which one would expect to be close to the measured energy using a
symmetric line shape, but the quantity is width-dependent ($E_{\mathrm{max}} =
E+W^{2}/E$), and \citet{muller_nocorrelation_2012} point out that the width of
the line can vary with choice of continuum model (e.g., for \nin,
\citet{makishima_1907_1999}, using NPEX, find a width of $\sim\!7$\,keV, while
\citet{inam_torque_2009} find widths of $\sim\!1-3$\,keV when they use PLCUT).
With this being the case, we opted to simply fit our data with a
pseudo-Lorentzian (the CYCLABS model in \textit{XSPEC}), obtaining values for
$E_{\mathrm{c1}}$ that were, on average, $1.75$\,keV lower than those we found
using GAUABS. This is roughly what one would expect using
\citet{mihara_thesis_1995}'s expression with a width of $\sim\!6$\,keV.

To investigate the correlation between luminosity and CRSF energy, we performed
linear fits and calculated 90\% error bars for the available references for
each source; however, due to the systematic uncertainty introduced by
correcting for the Lorentzian-modeled points, we defined two datasets for each
source: the first only taking into account those results that used Gaussian
line profiles similar to this work, the second with the ``corrected''
Lorentzian-modeled values as well as the Gaussian-modeled lines. The
historical data for each source is insufficient to draw any conclusions based
on the Lorentzian-modeled data alone. For \fif, the Gaussian-modeled data
amounts to eight points: the six values from Table \ref{tab:H1538_lum}, along
with measurements by \citet{robba_bepposax_2001} and
\citet{coburn_magnetic_2001}. A linear fit to these data finds a slope of $-3
\pm 18$\,\evslum, consistent with zero. Pearson's $r$ for this dataset is
-0.4; with six degrees of freedom, this is consistent with no correlation. If
we attempt to include the Lorentzian-modeled points in addition to the
Gaussian-modeled points, the size of the applied correction becomes a deciding
factor: any correction larger than $\sim 1.3$\,keV results in a negative slope
inconsistent with zero, with our +1.75\,keV shift resulting in a slope of $-7
\pm 4$\,\evslum. Pearson's $r$ with this correction applied is $0.13$, still
suggesting no correlation even with the additional data, although this is
tempered by the large error bars on several of the datapoints.
For \nin, the Gaussian-modeled dataset consists of this work along with the
results of \citet{cusumano_1907_1998}, \citet{coburn_magnetic_2001},
\citet{fritz_1907_2006}, and \citet{rivers_comprehensive_2010}. A linear fit to
these data produces a positive slope of $4 \pm 3$\,\evslum and a Pearson's $r$
of 0.92, indicative of a correlation at the $p = 0.02$ level. However,
we should note that this positive slope is dependent to some extent on the
measurements of \citet{rivers_comprehensive_2010} - specfically, the
higher-luminosity measurement of the CRSF. When that point is excluded from
the dataset, Pearson's r is still 0.93, but the slope is no longer
significant at the 90\% level, returning a value of $3^{+5}_{-6}$\,\evslum.
This weakens somewhat any conclusion that \nin definitely has any
CRSF-luminosity correlation. The discrepancy in the effect of correcting the
Lorentzian-modeled results of \citet{makishima_1907_1999} and
\citet{inam_torque_2009} makes the inclusion of these points more suspect than
was the case for \fif. Nonetheless, including the Lorentzian-modeled points,
regardless of whether they are ``corrected'' upwards or not, often still
results in a positive slope. Only when the shift is greater than $\sim
1.8$\,keV, or the \citet{rivers_comprehensive_2010} point is excluded
does the slope become consistent with zero.

Considering the possibility of a correlation between CRSF energy and
luminosity in \nin, it should be noted that a CRSF energy-luminosity correlation
at this low of a luminosity has not been seen until now. The only source of
comparable luminosity to \fif or \nin with a well-studied luminosity
\textit{vs.}\ cyclotron line energy relationship is A~0535+26
\citep{caballero_a0535_2007}, which shows no long-term trend.  A~0535+26's
positive trend is limited to the pulse-to-pulse analysis by
\citet{klochkov_pulse_2011}, which only covers higher-luminosity observations
where the source was above $L_{\mathrm{coul}}$ \footnote{The astute
reader may note that \citet{reig_bexrb_2013} also saw evidence for a positive
trend in RXTE/PCA data from A 0535+26's 2009 giant outburst.
\citet{nespoli_accretion_2013}, however, note that this trend was likely in
part due to an oversubtraction of the PCA background, reducing the
significance of the \citeauthor{reig_bexrb_2013} results.}. With \fif and
\nin operating at $L \lesssim L_{\mathrm{coul}}$, the potential presence of a
correlation is thus rather interesting - while the theoretical predictions
for this luminosity regime are still somewhat speculative, the general
concept was that at low luminosities, the scattering region would be
essentially at the surface of the star, and an increase in luminosity would
either do nothing, or lift the scattering region up off the surface of the
star, producing a negative trend in luminosity \textit{vs.}\ cyclotron line
energy. We see something like this in \fif, which potentially has a negative
slope in L-vs.-E space.  However, \nin shows evidence for a positive
correlation. This would make \nin the fourth CRSF source to display a positive
correlation with luminosity, after Hercules~X-1 \citep{staubert_herx1_2007},
A~0535+26 \citep{klochkov_pulse_2011}, and GX~304$-$1
\citep{yamamoto_gx301crsf_2011}.  Hercules~X-1 and GX~301$-$4 have L-vs.-E
slopes of similar magnitude to \nin, but they operate in a distinctly different
luminosity regime, at $L_{\mathrm{coul}} < L \lesssim L_{\mathrm{crit}}$.
The existence of a correlation in \nin would make the source somewhat
of an outlier, with its luminosity sitting as far below $L_{\mathrm{coul}}$
and $L_{\mathrm{crit}}$ as it is. However, the weakness of the correlation
and the low luminosity of the source make a conclusive determination difficult
with the quality of data available.

\section{Summary}
\label{sec:conclusion}
We have presented a spectral and timing analysis of the public \emph{INTEGRAL}
data for the high-mass X-ray binaries \fif and \nin. Our pulse period
measurements for \fif support the \emph{Fermi} GBM pulsar monitoring project's
observation of a switch to a spin-down trend starting sometime in 2009, with
our measured pulse periods agreeing strongly with their results. The source
shows no significant spectral changes between its spin-up and spin-down
epochs, and neither does its spectrum change much between higher and lower
luminosities. We update the spectral results for \nin to include the latest
public \emph{INTEGRAL} data, the results agreeing with much of the recent
work on the source. In both sources, we detect two absorption features,
identified as cyclotron scattering resonance features. The fundamental CRSF
in each source is detected very significantly at $\sim\!22$ and
$\sim\!19$\,keV in \fif and \nin, respectively, with harmonics detected in
\fif at $\sim\!50$\,keV and in \nin at $\sim\!38$\,keV. An examination of
our results along with those of past work reveals a possible negative
correlation between CRSF energy and luminosity in \fif, while \nin's CRSF
energy-luminosity relationship has evidence for a positive slope,
making it the fourth cyclotron line source to display this relationship.

\acknowledgements
\label{sec:acknowlegements}
PBH acknowleges support by National Aeronautics and Space Administration
grant NNX09AT26G.  IC acknowleges financial support from the French Space
Agency CNES though CNRS.  This work is based on observations made by the
\emph{INTEGRAL} satellite, a European Space Agency (ESA) project funded by
ESA member states.  We also thank the \emph{Fermi} GBM Pulsar Project for
their extended monitoring of multiple pulsars.

\bibliographystyle{apj}
\bibliography{hemp2013_integral}

\end{document}

%% file: H1538_table_date_rebinned.tex
\begin{deluxetable*}{llrrrr}
  \tabletypesize{\scriptsize}
  \tablecolumns{8}
  \tablewidth{0pt}
  \tablecaption{Spectral parameters for 4U~1538$-$522 for 2003-2008 and 2008-2010}
  \tablehead{ \colhead{} & 
              \colhead{} &
              \multicolumn{2}{c}{Early (2003-2008)} &
              \multicolumn{2}{c}{Late (2008-2010)} \\
              \multicolumn{2}{l}{Continuum} &
              \colhead{POWERLAW} &
              \colhead{CUTOFFPL+BUMP} &
              \colhead{POWERLAW} &
              \colhead{CUTOFFPL+BUMP}}
  \startdata
  $\Gamma$ &  & $1.17^{+0.06}_{-0.07}$ & $0.7^{+0.2}_{-0.2}$ & $1.12^{+0.08}_{-0.14}$ & $0.4^{+0.2}_{-0.5}$ \\
  Normalization & \tablenotemark{a} & $3.3^{+0.6}_{-0.5}$ & $2.2^{+0.8}_{-0.7}$ & $2.9^{+0.6}_{-0.7}$ & $1.6^{+0.5}_{-0.7}$ \\
  5-100\,keV flux & \tablenotemark{b} & $6.62^{+0.08}_{-3.36}$ & $6.48^{+0.14}_{-0.40}$ & $6.3^{+0.2}_{-0.6}$ & $6.3^{+0.2}_{-0.5}$\\
  $E_{\mathrm{cut}}$ & keV & $14.2^{+1.3}_{-1.0}$ & \nodata & $14^{+3}_{-2}$ & \nodata \\
  $E_{\mathrm{fold}}$ & keV & $14^{+4}_{-2}$ & $13^{+3}_{-2}$ & $11.7^{+1.3}_{-1.5}$ & $9.4^{+1.0}_{-1.6}$ \\
  $E_{\mathrm{bump}}$ & keV & \nodata & $13.0^{+0.8}_{-0.5}$ & \nodata & $13.3^{+1.0}_{-0.8}$ \\
  $\sigma_{\mathrm{bump}}$ & keV & \nodata & $1.7^{+0.8}_{-0.3}$ & \nodata & $1.70$ (frozen) \\
  $\mathrm{Intensity}_{\mathrm{bump}}$ & \tablenotemark{c} & \nodata & $1.4^{+1.3}_{-0.6}$ & \nodata & $1.2^{+2.4}_{-0.6}$ \\
  \hline \\[-1.5ex]
  $E_{\mathrm{c1}}$ & keV & $21.5^{+0.5}_{-0.6}$ & $22.0^{+0.6}_{-1.8}$ & $21.1^{+0.8}_{-0.8}$ & $21.9^{+0.8}_{-3.9}$ \\
  $\sigma_{\mathrm{c1}}$ & keV & $3.2^{+0.7}_{-0.6}$ & $2.9^{+1.9}_{-1.1}$ & $3.8^{+1.4}_{-1.2}$ & $3.4^{+5.0}_{-1.2}$ \\
  $\tau_{\mathrm{c1}}$ &  & $0.48^{+0.11}_{-0.07}$ & $0.32^{+0.09}_{-0.08}$ & $0.46^{+0.21}_{-0.13}$ & $0.38^{+0.11}_{-0.10}$ \\
  \hline \\[-1.5ex]
  $E_{\mathrm{c2}}$ & keV & $50^{+5}_{-4}$ & $50^{+4}_{-3}$ & \nodata & \nodata \\
  $\sigma_{\mathrm{c2}}$ & keV & $8^{+6}_{-3}$ & $8^{+3}_{-2}$ & \nodata & \nodata \\
  $\tau_{\mathrm{c2}}$ &  & $0.8^{+0.6}_{-0.3}$ & $1.0^{+0.4}_{-0.3}$ & \nodata & \nodata \\
  \hline \\[-1.5ex]
  JEM-X 1 normalization &  & $0.91^{+0.06}_{-0.06}$ & $0.96^{+0.06}_{-0.06}$ & $0.96^{+0.11}_{-0.10}$ & $0.92^{+0.09}_{-0.09}$ \\
  JEM-X 2 normalization &  & $1.02^{+0.07}_{-0.07}$ & $1.07^{+0.07}_{-0.07}$ & $1.07^{+0.12}_{-0.11}$ & $1.03^{+0.10}_{-0.10}$ \\
  \hline \\[-1.5ex]
  \multicolumn{2}{l}{$\chi^{2}_{\mathrm{red}}$ (dof)} & $1.175 (44)$ & $1.095 (43)$ & $0.7277 (46)$ & $0.6633 (45)$
  \enddata
  \tablenotetext{a}{Units: $10^{-2}$\,photons\,cm$^{-2}$\,s$^{-1}$ at 1 keV}
  \tablenotetext{b}{Units: $ 10^{-10}$\,erg\,cm$^{-2}$\,s$^{-1} $ }
  \tablenotetext{c}{Units: $10^{-3}$\,photons\,cm$^{-2}$\,s$^{-1}$}
  \label{tab:H1538_date}
\end{deluxetable*}

%% file: H1538_table_luminosity_rebinned.tex
\clearpage
\begin{turnpage}
  \begin{deluxetable}{llrrrrrrrrrrrr}
    \tabletypesize{\scriptsize}
    \setlength{\tabcolsep}{0.03in}
    \tablecolumns{14}
    \tablewidth{0pt}
    \tablecaption{Spectral parameters for 4U~1538$-$522 for three luminosity bins}
    \tablehead{ \colhead{} & \colhead{} &
                \multicolumn{4}{c}{High} &
                \multicolumn{4}{c}{Mid} &
                \multicolumn{4}{c}{Low} \\
                \multicolumn{2}{c}{ } &
                \multicolumn{2}{c}{Early} &
                \multicolumn{2}{c}{Late} &
                \multicolumn{2}{c}{Early} &
                \multicolumn{2}{c}{Late} &
                \multicolumn{2}{c}{Early} &
                \multicolumn{2}{c}{Late} \\
                \multicolumn{2}{c}{Continuum} &
                \colhead{P \tablenotemark{a}} &
                \colhead{C \tablenotemark{b}}&
                \colhead{P} &
                \colhead{C}&
                \colhead{P} &
                \colhead{CB \tablenotemark{c}}&
                \colhead{P} &
                \colhead{C}&
                \colhead{P} &
                \colhead{C}&
                \colhead{P} &
                \colhead{C}}
    \startdata
    $\Gamma$ &  & $1.07^{+0.10}_{-0.16}$ & $0.3^{+0.2}_{-0.2}$ & $1.04^{+0.12}_{-0.27}$ & $0.4^{+0.2}_{-0.3}$ & $1.17^{+0.07}_{-0.07}$ & $0.9^{+0.2}_{-0.2}$ & $1.06^{+0.09}_{-0.17}$ & $0.6^{+0.2}_{-0.2}$ & $1.20^{+0.12}_{-0.14}$ & $0.2^{+0.2}_{-0.2}$ & $1.18^{+0.12}_{-0.14}$ & $0.1^{+0.2}_{-0.2}$ \\
    Normalization & \tablenotemark{d} & $4.8^{+1.2}_{-1.3}$ & $2.1^{+0.6}_{-0.5}$ & $4.6^{+1.3}_{-1.6}$ & $2.4^{+0.9}_{-0.7}$ & $3.7^{+0.7}_{-0.6}$ & $3.6^{+1.5}_{-1.1}$ & $3.0^{+0.7}_{-0.4}$ & $1.8^{+0.7}_{-0.6}$ & $1.9^{+0.6}_{-0.5}$ & $0.6^{+0.3}_{-0.2}$ & $2.1^{+0.8}_{-0.6}$ & $0.7^{+0.3}_{-0.2}$ \\
    5-100\,keV flux & \tablenotemark{e} & $11.4^{+0.2}_{-1.4}$ & $11.4^{+0.2}_{-2.8}$ & $9.9^{+0.3}_{-2.4}$ & $10.1^{+0.2}_{-2.5}$ & $7.1^{+0.2}_{-0.4}$ & $7.1^{+0.2}_{-1.0}$ & $8^{+6}_{-2}$ & $8.3^{+0.3}_{-1.4}$ & $3.2^{+0.2}_{-0.4}$ & $3.22^{+0.07}_{-0.27}$ & $3.9^{+0.3}_{-0.4}$ & $4.0^{+0.2}_{-0.7}$ \\
    $E_{\mathrm{cut}}$ & keV & $14^{+2}_{-2}$ & \nodata & $13^{+4}_{-2}$ & \nodata & $13.8^{+1.8}_{-1.2}$ & \nodata & $13^{+2}_{-2}$ & \nodata & $22^{+5}_{-4}$ & \nodata & $14^{+7}_{-2}$ & \nodata \\
    $E_{\mathrm{fold}}$ & keV & $13^{+2}_{-2}$ & $11^{+2}_{-2}$ & $18^{+4}_{-4}$ & $15^{+3}_{-3}$ & $14^{+2}_{-2}$ & $14^{+4}_{-3}$ & $> 13.0$ & $14^{+4}_{-4}$ & $8^{+2}_{-2}$ & $8.2^{+1.0}_{-0.8}$ & $12^{+2}_{-4}$ & $8.1^{+1.3}_{-1.1}$ \\
    $E_{\mathrm{bump}}$ & keV & \nodata & \nodata & \nodata & \nodata & \nodata & $12.7^{+1.0}_{-0.6}$ & \nodata & \nodata & \nodata & \nodata & \nodata & \nodata \\
    $\sigma_{\mathrm{bump}}$ & keV & \nodata & \nodata & \nodata & \nodata & \nodata & $2.03$ (frozen) & \nodata & \nodata & \nodata & \nodata & \nodata & \nodata \\
    $\mathrm{Intensity}_{\mathrm{bump}}$ & \tablenotemark{f} & \nodata & \nodata & \nodata & \nodata & \nodata & $2.6^{+1.0}_{-0.9}$ & \nodata & \nodata & \nodata & \nodata & \nodata & \nodata \\
    $E_{\mathrm{c1}}$ & keV & $21.2^{+0.6}_{-0.7}$ & $22.0^{+0.5}_{-0.4}$ & $21.6^{+1.0}_{-1.1}$ & $21.9^{+0.9}_{-0.9}$ & $21.5^{+0.7}_{-0.7}$ & $22.1^{+0.7}_{-0.9}$ & $23.1^{+1.1}_{-1.3}$ & $24.1^{+1.5}_{-1.3}$ & $21.8^{+2.3}_{-1.4}$ & $21.3^{+0.8}_{-0.7}$ & $21.5^{+1.2}_{-1.1}$ & $21.9^{+1.2}_{-1.0}$ \\
    $\sigma_{\mathrm{c1}}$ & keV & $3.7^{+1.1}_{-1.1}$ & $3.4^{+0.7}_{-0.7}$ & $4.0^{+1.5}_{-1.2}$ & $3.8^{+1.2}_{-1.0}$ & $3.3^{+1.0}_{-0.9}$ & $2.8^{+2.1}_{-1.4}$ & $5.6^{+1.0}_{-1.6}$ & $5^{+2}_{-2}$ & $3.4^{+0.8}_{-0.9}$ & $2.2^{+1.1}_{-1.5}$ & $2^{+3}_{-2}$ & $1.6^{+1.6}_{-1.4}$ \\
    $\tau_{\mathrm{c1}}$ & & $0.56^{+0.22}_{-0.13}$ & $0.60^{+0.12}_{-0.13}$ & $0.6^{+0.2}_{-0.2}$ & $0.6^{+0.2}_{-0.2}$ & $0.45^{+0.13}_{-0.11}$ & $0.40^{+0.13}_{-0.11}$ & $0.6^{+0.2}_{-0.3}$ & $0.65^{+0.23}_{-0.10}$ & $1.0^{+0.3}_{-0.3}$ & $0.6^{+1.2}_{-0.2}$ & $0.8^{+7.9}_{-0.4}$ & $ >0.3$ \\
    $E_{\mathrm{c2}}$ & keV & $49^{+4}_{-3}$ & $49^{+4}_{-3}$ & $50^{+8}_{-4}$ & $49^{+7}_{-4}$ & $50^{+4}_{-3}$ & $53^{+4}_{-3}$ & $57^{+6}_{-13}$ & $53^{+4}_{-4}$ & \nodata & \nodata & \nodata & \nodata \\
    $\sigma_{\mathrm{c2}}$ & keV & $8^{+4}_{-3}$ & $10^{+3}_{-4}$ & $15^{+14}_{-4}$ & $16^{+11}_{-4}$ & $8$ (frozen) & $10$ (frozen) & $14^{+5}_{-11}$ & $10$ (frozen) & \nodata & \nodata & \nodata & \nodata \\
    $\tau_{\mathrm{c2}}$ & & $1.0^{+0.4}_{-0.4}$ & $1.1^{+0.4}_{-0.5}$ & $1.2^{+1.1}_{-0.5}$ & $1.5^{+1.1}_{-0.5}$ & $0.6^{+0.2}_{-0.3}$ & $1.0^{+0.4}_{-0.4}$ & $2.0^{+0.6}_{-1.5}$ & $1.8^{+1.4}_{-1.0}$ & \nodata & \nodata & \nodata & \nodata \\
    JEM-X 1 normalization & & $0.99^{+0.07}_{-0.07}$ & $1.01^{+0.10}_{-0.09}$ & $0.87^{+0.13}_{-0.12}$ & $0.89^{+0.13}_{-0.11}$ & $0.99^{+0.10}_{-0.09}$ & $0.93^{+0.08}_{-0.08}$ & $1.12^{+0.08}_{-0.13}$ & $1.14^{+0.14}_{-0.12}$ & $0.97^{+0.11}_{-0.10}$ & $0.94^{+0.11}_{-0.10}$ & $1.0^{+0.2}_{-0.2}$ & $1.1^{+0.2}_{-0.2}$ \\
    JEM-X 2 normalization & & $0.72^{+0.05}_{-0.05}$ & $0.73^{+0.07}_{-0.07}$ & $0.89^{+0.13}_{-0.13}$ & $0.91^{+0.13}_{-0.12}$ & $0.93^{+0.10}_{-0.09}$ & $0.87^{+0.08}_{-0.08}$ & $1.23^{+0.14}_{-0.14}$ & $1.25^{+0.15}_{-0.14}$ & $1.3^{+0.2}_{-0.2}$ & $1.31^{+0.16}_{-0.15}$ & $1.1^{+0.2}_{-0.2}$ & $1.1^{+0.2}_{-0.2}$ \\
    \hline
    $\chi^{2}_{\mathrm{red}}$ (dof) & & $1.038 (54)$ & $1.134 (55)$ & $0.7066 (50)$ & $0.7644 (51)$ & $1.060 (46)$ & $1.090 (45)$ & $0.8842 (50)$ & $0.9487 (52)$ & $1.224 (41)$ & $1.356 (42)$ & $0.4921 (41)$ & $0.5870 (42)$
    \enddata\\[-2ex]
    \tablenotetext{a}{PLCUT continuum\\[-1ex]}
    \tablenotetext{b}{CUTOFFPL continuum\\[-1ex]}
    \tablenotetext{c}{CUTOFFPL+BUMP continuum\\[-1ex]}
    \tablenotetext{d}{Units: $10^{-2}$\,photons\,cm$^{-2}$\,s$^{-1}$ at 1 keV\\[-1ex]}
    \tablenotetext{e}{Units: $ 10^{-10}$\,erg\,cm$^{-2}$\,s$^{-1} $ \\[-1ex]}
    \tablenotetext{f}{Units: $10^{-3}$\,photons\,cm$^{-2}$\,s$^{-1}$\\[-1ex]}
    \label{tab:H1538_lum}
  \end{deluxetable}
\clearpage
\end{turnpage}

%% file: 4u1907_table_rebinned.tex
\begin{deluxetable}{llrr}
  \tabletypesize{\scriptsize}
  \tablecolumns{4}
  \tablewidth{0pt}
  \tablecaption{Spectral parameters for 4U~1907$+$09 \label{tab:4U1907_full_spectrum}}
  \tablehead{ \multicolumn{2}{l}{Continuum} &
              \colhead{POWERLAW} &
              \colhead{CUTOFFPL+BUMP}}
  \startdata
  $\Gamma$ &  & $-0.9^{+0.8}_{-0.8}$ & $-1.3^{+1.0}_{-0.7}$ \\
  Normalization & \tablenotemark{a} & $0.03^{+0.09}_{-0.02}$ & $0.04^{+0.1}_{-0.03}$ \\
  5-100\,keV flux & \tablenotemark{b} & $2.92^{+0.04}_{-0.31}$ & $2.92^{+0.05}_{-0.20}$ \\
  $E_{\mathrm{cut}}$ & keV & $6.5^{+1.6}_{-2.4}$ & \nodata \\
  $E_{\mathrm{fold}}$ & keV & $6.2^{+3.8}_{-1.3}$ & $5.6^{+17.2}_{-1.2}$ \\
  \hline \\[-1.5ex]
  $E_{\mathrm{c1}}$ & keV & $18.5^{+0.7}_{-0.7}$ & $18.4^{+0.7}_{-0.7}$ \\
  $\sigma_{\mathrm{c1}}$ & keV & $3.2^{+1.1}_{-0.8}$ & $3.5^{+1.1}_{-0.8}$ \\
  $\tau_{\mathrm{c1}}$ &  & $0.8^{+0.2}_{-0.2}$ & $0.9^{+0.3}_{-0.2}$ \\
  \hline \\[-1.5ex]
  $E_{\mathrm{c2}}$ & keV & $38^{+8}_{-5}$ & $38^{+14}_{-5}$ \\
  $\sigma_{\mathrm{c2}}$ & keV & $8^{+14}_{-6}$ & $9^{+18}_{-5}$ \\
  $\tau_{\mathrm{c2}}$ &  & $1.1^{+5.4}_{-0.8}$ & $1.2^{+5.1}_{-0.9}$ \\
  \hline \\[-1.5ex]
  JEM-X 1 normalization &  & $0.95^{+0.13}_{-0.12}$ & $0.95^{+0.13}_{-0.12}$ \\
  JEM-X 2 normalization &  & $1.19^{+0.16}_{-0.15}$ & $1.19^{+0.16}_{-0.15}$ \\
  \hline \\[-1.5ex]
  \multicolumn{2}{l}{$\chi^{2}_{\mathrm{red}}$ (dof)} & $0.7774 (35)$ & $0.8260 (36)$ 
  \enddata
  \tablenotetext{a}{Units: $10^{-2}$\,photons\,cm$^{-2}$\,s$^{-1}$ at 1 keV}
  \tablenotetext{b}{Units: $ 10^{-10}$\,erg\,cm$^{-2}$\,s$^{-1} $ }
  \tablenotetext{c}{Units: $10^{-3}$\,photons\,cm$^{-2}$\,s$^{-1}$}
\end{deluxetable}

%% file: hemp2013_integral.bbl
\begin{thebibliography}{64}
\expandafter\ifx\csname natexlab\endcsname\relax\def\natexlab#1{#1}\fi

\bibitem[{{Arnaud}(1996)}]{arnaud_xspec_1996}
{Arnaud}, K.~A. 1996, in Astronomical Society of the Pacific Conference Series,
  Vol. 101, Astronomical Data Analysis Software and Systems V, ed. G.~H.
  {Jacoby} \& J.~{Barnes}, 17

\bibitem[{Baykal {et~al.}(2006)Baykal, \.{I}nam, \&
  Beklen}]{baykal_recent_2006}
Baykal, A., \.{I}nam, S.~c., \& Beklen, E. 2006, \aap, 453, 1037

\bibitem[{{Becker} \& {Wolff}(2007)}]{becker_continuum_2007}
{Becker}, P.~A., \& {Wolff}, M.~T. 2007, \apj, 654, 435

\bibitem[{{Becker} {et~al.}(2012){Becker}, {Klochkov}, {Sch{\"o}nherr},
  {Nishimura}, {Ferrigno}, {Caballero}, {Kretschmar}, {Wolff}, {Wilms}, \&
  {Staubert}}]{becker_spectral_2012}
{Becker}, P.~A., {Klochkov}, D., {Sch{\"o}nherr}, G., {et~al.} 2012, \aap, 544,
  A123

\bibitem[{{Becker} {et~al.}(1977){Becker}, {Swank}, {Boldt}, {Holt},
  {Serlemitsos}, {Pravdo}, \& {Saba}}]{becker_1538_1977}
{Becker}, R.~H., {Swank}, J.~H., {Boldt}, E.~A., {et~al.} 1977, \apjl, 216, L11

\bibitem[{{{\c S}ahiner} {et~al.}(2011){{\c S}ahiner}, {Inam}, \&
  {Baykal}}]{sahiner_recent_2011}
{{\c S}ahiner}, {\c S}., {Inam}, S.~{\c C}., \& {Baykal}, A. 2011, in American
  Institute of Physics Conference Series, Vol. 1379, American Institute of
  Physics Conference Series, ed. E.~{G{\"o}{\u g}{\"u}{\c s}}, T.~{Belloni}, \&
  {\"U}.~{Ertan}, 214--216

\bibitem[{{{\c S}ahiner} {et~al.}(2012){{\c S}ahiner}, {Inam}, \&
  {Baykal}}]{sahiner_comprehensive_2012}
{{\c S}ahiner}, {\c S}., {Inam}, S.~{\c C}., \& {Baykal}, A. 2012, \mnras, 421,
  2079

\bibitem[{{Caballero} {et~al.}(2007){Caballero}, {Kretschmar}, {Santangelo},
  {Staubert}, {Klochkov}, {Camero}, {Ferrigno}, {Finger}, {Kreykenbohm},
  {McBride}, {Pottschmidt}, {Rothschild}, {Sch{\"o}nherr}, {Segreto}, {Suchy},
  {Wilms}, \& {Wilson}}]{caballero_a0535_2007}
{Caballero}, I., {Kretschmar}, P., {Santangelo}, A., {et~al.} 2007, \aap, 465,
  L21

\bibitem[{{Camero-Arranz} {et~al.}(2010){Camero-Arranz}, {Finger}, {Ikhsanov},
  {Wilson-Hodge}, \& {Beklen}}]{cameroarranz_torque1626_2010}
{Camero-Arranz}, A., {Finger}, M.~H., {Ikhsanov}, N.~R., {Wilson-Hodge}, C.~A.,
  \& {Beklen}, E. 2010, \apj, 708, 1500

\bibitem[{{Chernyakova} {et~al.}(2010){Chernyakova}, {Neronov}, {Walter}, \&
  {Courvoisier}}]{osa_ibis}
{Chernyakova}, M., {Neronov}, A., {Walter}, R., \& {Courvoisier}, T. 2010, IBIS
  Analysis User Manual, INTEGRAL Science Data Centre

\bibitem[{Clark(2000)}]{clark_orbit_2000}
Clark, G.~W. 2000, \apj, 542, L131

\bibitem[{{Clark}(2004)}]{clark_chandra1538_2004}
{Clark}, G.~W. 2004, \apj, 610, 956

\bibitem[{Clark {et~al.}(1990)Clark, Woo, Nagase, Makishima, \&
  Sakao}]{clark_discovery_1990}
Clark, G.~W., Woo, J.~W., Nagase, F., Makishima, K., \& Sakao, T. 1990, \apj,
  353, 274

\bibitem[{Coburn(2001)}]{coburn_magnetic_2001}
Coburn, W. 2001, Phd, University of California, San Diego

\bibitem[{{Corbet} {et~al.}(1993){Corbet}, {Woo}, \&
  {Nagase}}]{corbet_orbit_1993}
{Corbet}, R.~H.~D., {Woo}, J.~W., \& {Nagase}, F. 1993, \aap, 276, 52

\bibitem[{{Cox} {et~al.}(2005){Cox}, {Kaper}, \& {Mokiem}}]{cox_1907_2005}
{Cox}, N.~L.~J., {Kaper}, L., \& {Mokiem}, M.~R. 2005, \aap, 436, 661

\bibitem[{{Crampton} {et~al.}(1978){Crampton}, {Hutchings}, \&
  {Cowley}}]{crampton_1538_1978}
{Crampton}, D., {Hutchings}, J.~B., \& {Cowley}, A.~P. 1978, \apjl, 225, L63

\bibitem[{{Cusumano} {et~al.}(1998){Cusumano}, {di Salvo}, {Burderi},
  {Orlandini}, {Piraino}, {Robba}, \& {Santangelo}}]{cusumano_1907_1998}
{Cusumano}, G., {di Salvo}, T., {Burderi}, L., {et~al.} 1998, \aap, 338, L79

\bibitem[{{Davison}(1977)}]{davison_1538_1977a}
{Davison}, P.~J.~N. 1977, \mnras, 179, 35P

\bibitem[{{Davison} {et~al.}(1977){Davison}, {Watson}, \&
  {Pye}}]{davison_1538_1977b}
{Davison}, P.~J.~N., {Watson}, M.~G., \& {Pye}, J.~P. 1977, \mnras, 181, 73P

\bibitem[{{Finger} {et~al.}(2009){Finger}, {Beklen}, {Narayana Bhat},
  {Paciesas}, {Connaughton}, {Buckley}, {Camero-Arranz}, {Coe}, {Jenke},
  {Kanbach}, {Negueruela}, \& {Wilson-Hodge}}]{finger_gbm_2009}
{Finger}, M.~H., {Beklen}, E., {Narayana Bhat}, P., {et~al.} 2009, in 2009
  Fermi Symposium

\bibitem[{{Fritz} {et~al.}(2006){Fritz}, {Kreykenbohm}, {Wilms}, {Staubert},
  {Bayazit}, {Pottschmidt}, {Rodriguez}, \& {Santangelo}}]{fritz_1907_2006}
{Fritz}, S., {Kreykenbohm}, I., {Wilms}, J., {et~al.} 2006, \aap, 458, 885

\bibitem[{Giacconi {et~al.}(1971)Giacconi, Kellogg, Gorenstein, Gursky, \&
  Tananbaum}]{giacconi_x-ray_1971}
Giacconi, R., Kellogg, E., Gorenstein, P., Gursky, H., \& Tananbaum, H. 1971,
  \apj, 165, L27

\bibitem[{Giacconi {et~al.}(1974)Giacconi, Murray, Gursky, Kellogg, Schreier,
  Matilsky, Koch, \& Tananbaum}]{giacconi_third_1974}
Giacconi, R., Murray, S., Gursky, H., {et~al.} 1974, \apjs, 27, 37

\bibitem[{{Ilovaisky} {et~al.}(1979){Ilovaisky}, {Chevalier}, \&
  {Motch}}]{ilovaisky_1538_1979}
{Ilovaisky}, S.~A., {Chevalier}, C., \& {Motch}, C. 1979, \aap, 71, L17

\bibitem[{{in 't Zand} {et~al.}(1998){in 't Zand}, {Baykal}, \&
  {Strohmayer}}]{intzand_1907_1998}
{in 't Zand}, J.~J.~M., {Baykal}, A., \& {Strohmayer}, T.~E. 1998, \apj, 496,
  386

\bibitem[{{Inam} {et~al.}(2009){Inam}, {{\c S}ahiner}, \&
  {Baykal}}]{inam_torque_2009}
{Inam}, S.~{\c C}., {{\c S}ahiner}, {\c S}., \& {Baykal}, A. 2009, \mnras, 395,
  1015

\bibitem[{{Klein} {et~al.}(1996){Klein}, {Arons}, {Jernigan}, \&
  {Hsu}}]{klein_accretion_1996}
{Klein}, R.~I., {Arons}, J., {Jernigan}, G., \& {Hsu}, J.~J.-L. 1996, \apjl,
  457, L85

\bibitem[{{Klochkov} {et~al.}(2011){Klochkov}, {Staubert}, {Santangelo},
  {Rothschild}, \& {Ferrigno}}]{klochkov_pulse_2011}
{Klochkov}, D., {Staubert}, R., {Santangelo}, A., {Rothschild}, R.~E., \&
  {Ferrigno}, C. 2011, \aap, 532, A126

\bibitem[{Kostka \& Leahy(2010)}]{kostka_evidence_2010}
Kostka, M., \& Leahy, D.~A. 2010, \mnras, 407, 1182

\bibitem[{{Lamb} {et~al.}(1973){Lamb}, {Pethick}, \&
  {Pines}}]{lamb_accretion_1973}
{Lamb}, F.~K., {Pethick}, C.~J., \& {Pines}, D. 1973, \apj, 184, 271

\bibitem[{{Larsson}(1996)}]{larsson_epfold_1996}
{Larsson}, S. 1996, \aaps, 117, 197

\bibitem[{{Leahy} {et~al.}(1983){Leahy}, {Darbro}, {Elsner}, {Weisskopf},
  {Kahn}, {Sutherland}, \& {Grindlay}}]{leahy_epfold_1983}
{Leahy}, D.~A., {Darbro}, W., {Elsner}, R.~F., {et~al.} 1983, \apj, 266, 160

\bibitem[{{Lebrun} {et~al.}(2003){Lebrun}, {Leray}, {Lavocat}, {Cr{\'e}tolle},
  {Arqu{\`e}s}, {Blondel}, {Bonnin}, {Bou{\`e}re}, {Cara}, {Chaleil}, {Daly},
  {Desages}, {Dzitko}, {Horeau}, {Laurent}, {Limousin}, {Mathy}, {Mauguen},
  {Meignier}, {Molini{\'e}}, {Poindron}, {Rouger}, {Sauvageon}, \&
  {Tourrette}}]{lebrun_isgri_2003}
{Lebrun}, F., {Leray}, J.~P., {Lavocat}, P., {et~al.} 2003, \aap, 411, L141

\bibitem[{{Lund} {et~al.}(2003){Lund}, {Budtz-J{\o}rgensen}, {Westergaard},
  {Brandt}, {Rasmussen}, {Hornstrup}, {Oxborrow}, {Chenevez}, {Jensen},
  {Laursen}, {Andersen}, {Mogensen}, {Rasmussen}, {Om{\o}}, {Pedersen},
  {Polny}, {Andersson}, {Andersson}, {K{\"a}m{\"a}r{\"a}inen}, {Vilhu},
  {Huovelin}, {Maisala}, {Morawski}, {Juchnikowski}, {Costa}, {Feroci},
  {Rubini}, {Rapisarda}, {Morelli}, {Carassiti}, {Frontera}, {Pelliciari},
  {Loffredo}, {Mart{\'{\i}}nez N{\'u}{\~n}ez}, {Reglero}, {Velasco}, {Larsson},
  {Svensson}, {Zdziarski}, {Castro-Tirado}, {Attina}, {Goria}, {Giulianelli},
  {Cordero}, {Rezazad}, {Schmidt}, {Carli}, {Gomez}, {Jensen}, {Sarri},
  {Tiemon}, {Orr}, {Much}, {Kretschmar}, \& {Schnopper}}]{lund_jemx_2003}
{Lund}, N., {Budtz-J{\o}rgensen}, C., {Westergaard}, N.~J., {et~al.} 2003,
  \aap, 411, L231

\bibitem[{{Makishima} {et~al.}(1984){Makishima}, {Kawai}, {Koyama},
  {Shibazaki}, {Nagase}, \& {Nakagawa}}]{makishima_1907_1984}
{Makishima}, K., {Kawai}, N., {Koyama}, K., {et~al.} 1984, \pasj, 36, 679

\bibitem[{{Makishima} {et~al.}(1987){Makishima}, {Koyama}, {Hayakawa}, \&
  {Nagase}}]{makishima_spectra_1987}
{Makishima}, K., {Koyama}, K., {Hayakawa}, S., \& {Nagase}, F. 1987, \apj, 314,
  619

\bibitem[{{Makishima} {et~al.}(1999){Makishima}, {Mihara}, {Nagase}, \&
  {Tanaka}}]{makishima_1907_1999}
{Makishima}, K., {Mihara}, T., {Nagase}, F., \& {Tanaka}, Y. 1999, \apj, 525,
  978

\bibitem[{{Marshall} \& {Ricketts}(1980)}]{marshall_1907_1980}
{Marshall}, N., \& {Ricketts}, M.~J. 1980, \mnras, 193, 7P

\bibitem[{{Meszaros} \& {Nagel}(1985)}]{meszaros_comptonization_1985b}
{Meszaros}, P., \& {Nagel}, W. 1985, \apj, 299, 138

\bibitem[{{Mihara}(1995)}]{mihara_thesis_1995}
{Mihara}, T. 1995, PhD thesis, Dept.~of Physics, Univ.~of Tokyo

\bibitem[{{Mihara} {et~al.}(1998){Mihara}, {Makishima}, \&
  {Nagase}}]{mihara_cyclovar_1998}
{Mihara}, T., {Makishima}, K., \& {Nagase}, F. 1998, Advances in Space
  Research, 22, 987

\bibitem[{Mukherjee {et~al.}(2006)Mukherjee, Raichur, Paul, Naik, \&
  Bhatt}]{mukherjee_orbital_2006}
Mukherjee, U., Raichur, H., Paul, B., Naik, S., \& Bhatt, N. 2006, Journal of
  Astrophysics and Astronomy, 27, 411

\bibitem[{{M{\"u}ller} {et~al.}(2012{\natexlab{a}}){M{\"u}ller}, {Ferrigno},
  {K{\"u}hnel}, {Sch{\"o}nherr}, {Becker}, {Wolff}, {Hertel}, {Schwarm},
  {Grinberg}, {Obst}, {Caballero}, {Pottschmidt}, {F{\"u}rst}, {Kreykenbohm},
  {Rothschild}, {Hemphill}, {Mart{\'{\i}}nez N{\'u}{\~n}ez}, {Torrej{\'o}n},
  {Klochkov}, {Staubert}, \& {Wilms}}]{muller_nocorrelation_2012}
{M{\"u}ller}, S., {Ferrigno}, C., {K{\"u}hnel}, M., {et~al.}
  2012{\natexlab{a}}, \aap, in press

\bibitem[{{M{\"u}ller} {et~al.}(2012{\natexlab{b}}){M{\"u}ller}, {K{\"u}hnel},
  {Caballero}, {Pottschmidt}, {F{\"u}rst}, {Kreykenbohm}, {Sagredo}, {Obst},
  {Wilms}, {Ferrigno}, {Rothschild}, \&
  {Staubert}}]{muller_sleeping_J1946_2012}
{M{\"u}ller}, S., {K{\"u}hnel}, M., {Caballero}, I., {et~al.}
  2012{\natexlab{b}}, \aap, 546, A125

\bibitem[{{Nespoli} {et~al.}(2008){Nespoli}, {Fabregat}, \&
  {Mennickent}}]{nespoli_1907_2008}
{Nespoli}, E., {Fabregat}, J., \& {Mennickent}, R.~E. 2008, \aap, 486, 911

\bibitem[{{Nespoli} {et~al.}(2013){Nespoli}, {Klochkov}, {Caballero}, {Reig},
  \& {Kretschmar}}]{nespoli_accretion_2013}
{Nespoli}, E., {Klochkov}, D., {Caballero}, I., {Reig}, P., \& {Kretschmar}, P.
  2013, in Spectral/Timing Properties of accreting objects: from X-ray binaries
  to AGN, ESA/ESAC, Madrid, Spain

\bibitem[{{Perna} {et~al.}(2006){Perna}, {Bozzo}, \&
  {Stella}}]{perna_spin_2006}
{Perna}, R., {Bozzo}, E., \& {Stella}, L. 2006, \apj, 639, 363

\bibitem[{{Rawls} {et~al.}(2011){Rawls}, {Orosz}, {McClintock}, {Torres},
  {Bailyn}, \& {Buxton}}]{rawls_mass_2011}
{Rawls}, M.~L., {Orosz}, J.~A., {McClintock}, J.~E., {et~al.} 2011, \apj, 730,
  25

\bibitem[{{Reig} \& {Nespoli}(2013)}]{reig_bexrb_2013}
{Reig}, P., \& {Nespoli}, E. 2013, \aap, 551, A1

\bibitem[{{Reynolds} {et~al.}(1992){Reynolds}, {Bell}, \&
  {Hilditch}}]{reynolds_optical_1538_1992}
{Reynolds}, A.~P., {Bell}, S.~A., \& {Hilditch}, R.~W. 1992, \mnras, 256, 631

\bibitem[{{Rivers} {et~al.}(2010){Rivers}, {Markowitz}, {Pottschmidt}, {Roth},
  {Barrag{\'a}n}, {F{\"u}rst}, {Suchy}, {Kreykenbohm}, {Wilms}, \&
  {Rothschild}}]{rivers_comprehensive_2010}
{Rivers}, E., {Markowitz}, A., {Pottschmidt}, K., {et~al.} 2010, \apj, 709, 179

\bibitem[{{Robba} {et~al.}(2001){Robba}, {Burderi}, {Di Salvo}, {Iaria}, \&
  {Cusumano}}]{robba_bepposax_2001}
{Robba}, N.~R., {Burderi}, L., {Di Salvo}, T., {Iaria}, R., \& {Cusumano}, G.
  2001, \apj, 562, 950

\bibitem[{{Rodes-Roca} {et~al.}(2010){Rodes-Roca}, Page, Torrej\'{o}n, Osborne,
  \& Bernab\'{e}u}]{rodes-roca_detecting_2010}
{Rodes-Roca}, J.~J., Page, K.~L., Torrej\'{o}n, J.~M., Osborne, J.~P., \&
  Bernab\'{e}u, G. 2010, \aap, 526, A64

\bibitem[{{Rodes-Roca} {et~al.}(2009){Rodes-Roca}, Torrej\'{o}n, Kreykenbohm,
  Mart\'{i}nez N\'{u}\~{n}ez, {Camero-Arranz}, \&
  Bernab\'{e}u}]{rodes-roca_first_2009}
{Rodes-Roca}, J.~J., Torrej\'{o}n, J.~M., Kreykenbohm, I., {et~al.} 2009, \aap,
  508, 395

\bibitem[{Rubin {et~al.}(1997)Rubin, Finger, Scott, \&
  Wilson}]{rubin_observation_1997}
Rubin, B.~C., Finger, M.~H., Scott, D.~M., \& Wilson, R.~B. 1997, \apj, 488,
  413

\bibitem[{{Sch{\"o}nherr} {et~al.}(2007){Sch{\"o}nherr}, {Wilms}, {Kretschmar},
  {Kreykenbohm}, {Santangelo}, {Rothschild}, {Coburn}, \&
  {Staubert}}]{schoenherr_CRSF_2007}
{Sch{\"o}nherr}, G., {Wilms}, J., {Kretschmar}, P., {et~al.} 2007, \aap, 472,
  353

\bibitem[{{Schwarm}(2013)}]{schwarm_thesis_2013}
{Schwarm}, F. 2013, Phd, University of {Erlangen-Nuremberg}

\bibitem[{{Schwarm} {et~al.}(2013){Schwarm}, {Sch{\"o}nherr}, {Becker},
  {Wolff}, {Wilms}, {Ferrigno}, \& {West}}]{schwarm_headposter_2013}
{Schwarm}, F.-W., {Sch{\"o}nherr}, G., {Becker}, P.~A., {et~al.} 2013, in
  AAS/High Energy Astrophysics Division, Vol.~13, AAS/High Energy Astrophysics
  Division

\bibitem[{{Staubert} {et~al.}(2007){Staubert}, {Shakura}, {Postnov}, {Wilms},
  {Rothschild}, {Coburn}, {Rodina}, \& {Klochkov}}]{staubert_herx1_2007}
{Staubert}, R., {Shakura}, N.~I., {Postnov}, K., {et~al.} 2007, \aap, 465, L25

\bibitem[{{Ubertini} {et~al.}(2003){Ubertini}, {Lebrun}, {Di Cocco}, {Bazzano},
  {Bird}, {Broenstad}, {Goldwurm}, {La Rosa}, {Labanti}, {Laurent}, {Mirabel},
  {Quadrini}, {Ramsey}, {Reglero}, {Sabau}, {Sacco}, {Staubert}, {Vigroux},
  {Weisskopf}, \& {Zdziarski}}]{ubertini_ibis_2003}
{Ubertini}, P., {Lebrun}, F., {Di Cocco}, G., {et~al.} 2003, \aap, 411, L131

\bibitem[{{van Kerkwijk} {et~al.}(1989){van Kerkwijk}, {van Oijen}, \& {van den
  Heuvel}}]{vankerkwijk_1907_1989}
{van Kerkwijk}, M.~H., {van Oijen}, J.~G.~J., \& {van den Heuvel}, E.~P.~J.
  1989, \aap, 209, 173

\bibitem[{{van Kerkwijk} {et~al.}(1995){van Kerkwijk}, {van Paradijs}, \&
  {Zuiderwijk}}]{vankerkwijk_masses_1995}
{van Kerkwijk}, M.~H., {van Paradijs}, J., \& {Zuiderwijk}, E.~J. 1995, \aap,
  303, 497

\bibitem[{{Yamamoto} {et~al.}(2011){Yamamoto}, {Sugizaki}, {Mihara},
  {Nakajima}, {Yamaoka}, {Matsuoka}, {Morii}, \&
  {Makishima}}]{yamamoto_gx301crsf_2011}
{Yamamoto}, T., {Sugizaki}, M., {Mihara}, T., {et~al.} 2011, \pasj, 63, 751

\end{thebibliography}
